\documentclass{article}
\pdfoutput=1

\usepackage[utf8]{inputenc}

\usepackage{setspace}   
\usepackage{lmodern} 
\usepackage{comment} 

\usepackage{amsmath}
\usepackage{nicefrac}
\usepackage{mathbbol}
\usepackage{booktabs}
\usepackage{adjustbox}
\usepackage{hyperref}
\usepackage[super]{nth}
\usepackage{graphicx}       
\usepackage{subcaption}
\usepackage[dvipsnames]{xcolor}

\usepackage[acronym,section,nonumberlist]{glossaries}
\newacronym{irr}{IRR}{Inter-Rater Reliability}

\newcommand{\given}{\, | \,}
\usepackage[style=apa,sortcites=true,sorting=nyt,backend=biber]{biblatex}
\DeclareLanguageMapping{american}{american-apa}
\newcommand{\beginappendix}{%
        \setcounter{table}{0}
        \renewcommand{\thetable}{A\arabic{table}}%
        \setcounter{figure}{0}
        \renewcommand{\thefigure}{A\arabic{figure}}%
     }
     
\usepackage{arxiv}

\title{Assessing inter-rater reliability with heterogeneous variance components models: Flexible approach accounting for contextual variables} \renewcommand{\shorttitle}{Assessing inter-rater reliability with heterogeneous variance components models}
\date{}
\author{Patrícia Martinková$^{1,2}$, František Bartoš$^{1,3}$, Marek Brabec$^{1}$\\
\small $^{1}$ Institute of Computer Science of the Czech Academy of Sciences, Prague, Czech Republic\\
\small $^{2}$ Faculty of Education, Charles University, Prague, Czech Republic\\
\small $^{3}$ Department of Psychological Methods, University of Amsterdam, Amsterdam, Netherlands.
}
\begin{document}

\maketitle

\thispagestyle{empty}


\begin{abstract}

Inter-rater reliability (IRR), which is a prerequisite of high-quality ratings and assessments, may be affected by contextual variables such as the rater’s or ratee’s gender, major, or experience. Identification of such heterogeneity sources in IRR is important for implementation of policies with the potential to decrease measurement error and to increase IRR by focusing on the most relevant subgroups. In this study, we propose a flexible approach for assessing IRR in cases of heterogeneity due to covariates by directly modeling differences in variance components. We use Bayes factors to select the best performing model, and we suggest using Bayesian model-averaging as an alternative approach for obtaining IRR and variance component estimates, allowing us to account for model uncertainty. We use inclusion Bayes factors considering the whole model space to provide evidence for or against differences in variance components due to covariates. The proposed method is compared with other Bayesian and frequentist approaches in a simulation study, and we demonstrate its superiority in some situations. Finally, we provide real data examples from grant proposal peer-review, demonstrating the usefulness of this method and its flexibility in the generalization of more complex designs. 
\end{abstract}
Keywords: Bayesian inference; inter-rater reliability; mixed-effect models; heterogeneous variance components; grant peer review

\newpage

\clearpage
\pagenumbering{arabic} 

\section{Introduction}
\vspace{-1.1em}
Inter-rater reliability (IRR) has been used to assess the quality of ratings and assessments in psychology, education, health, hiring, proposal and journal peer review, and with other areas involving multiple raters. From a measurement perspective, individual ratings (such as scores applicants receive from a hiring committee)
may be thought of as imprecise estimates of the true underlying quality of a measured subject or object. IRR enumerates the consistency among raters, and it may be described as the correlation between scores of different raters given to the same subject or object of measurement \parencite{webb2006}. 

A notable portion of research is focused on the identification of heterogeneity sources in IRR with respect to contextual variables, such as rater or ratee characteristics, with the goal of identifying policies with the potential to generally decrease measurement error and to increase the IRR especially for the lower-IRR subgroups. For example, 
IRR  was found to vary for different research areas of grant-proposal peer review \parencite{mutz2012}, and to increase after reviewer training \parencite{sattler2015grant}. In the context of teacher hiring, IRR was found to be lower for internal than external applicants  \parencite{martinkova2018disparities}, and lower for novice than experienced applicants \parencite{goldhaber2021profreferences}. In the context of teacher assessment, the IRR was found to be higher for the ratings of live vs. recorded lectures \parencite{casabianca2013effect}.

Different estimation techniques were considered in the past to account for heterogeneity in IRR with respect to groups. The most common approach involves stratification of the data and separate estimation of IRR in subgroups with ANOVA or mixed-effect models \parencite{sattler2015grant}. 
More complex mixed-effect models allowing for heterogeneous variance components were shown to detect group differences in IRR even in  those cases where no difference was detected using the stratification approach \parencite{martinkova2018disparities}. Another method is based upon generalized-estimation equations \parencite[GEE,][]{mutz2012}. \textcite{bartos2020testing} compared different methods for IRR estimation under heterogeneity with respect to a single grouping variable in a simulation study where the data-generating model was known, showing that both frequentist and Bayesian mixed-effect models, as well as general additive models, can provide accurate estimates of group-dependent IRR. 

However, further methodological complexities arise in real-life situations which were not solved by previous studies. The variance components of the ratings may be affected by a combination of contextual factors, such as the rater’s or ratee’s age, gender, major, or internal vs. external status, which may be of different types (besides binary also nominal, ordinal, or metric).  Furthermore, the model specification --- inclusion or exclusion of the different contextual factors ---  might need to be inferred from the data. Researchers might also be interested in whether a particular contextual factor does or does not affect the variance components, i.e., testing a hypothesis whether the effect of a given factor differs from zero. To our best knowledge, there has been no study estimating IRR or reliability with heterogeneous variance mixed-effects models in cases of heterogeneity due to a combination of covariates. None have dealt with model selection, nor is any general approach available. 

To fill this existing research gap, we propose a flexible general approach to IRR estimation and hypothesis testing using Bayes factors (BF) in those cases of variance heterogeneity due to covariates and unknown data-generating models. Our work builds upon studies of Bayesian 
mixed-effects models with heterogeneous variance components \parencite{williams2019bayesian, williams2020beneath} which were previously shown to provide a richer understanding of the psychological processes in various contexts, and upon the work of \textcite{dablander2020default} who recently introduced a default Bayes factor test for the inequality of variances. 
We also consider model-averaged estimates which incorporate uncertainty in the model selection process for the final estimate \parencite{hoeting1999bayesian, depaoli2020bayesian, raftery1995accounting, raftery1996approximate}. 

The paper proceeds as follows: We first introduce the  IRR in the multilevel modeling framework. We extend it to heterogeneous variance components, and introduce Bayesian hypothesis testing and model-averaging in Section~\ref{sec:Methods}. Second, we describe a simulation study and compare the proposed methodology to alternative approaches in Section~\ref{sec:Simulation}. Third, we illustrate the methodology on real data sets from ratings of the grant proposals in Section~\ref{sec:RealDataExample}. Finally, in  Section~\ref{sec:Discussion}, we conclude with a discussion of the results, and of further computational aspects of IRR estimation, including the aspects of generalizations in more complex designs. Sample \texttt{R} code and additional tables and figures are provided in electronic Supplementary Material at \href{https://osf.io/bk8a7/}{https://osf.io/bk8a7/}.

\section{Methods}\label{sec:Methods}

\subsection{IRR in multilevel modeling framework}
\label{sec:meth:IRR}

In the simplest case of a multilevel linear model with a one-way analysis of variance, rating $j$ of subject $i$, denoted $Y_{ij}$, is modelled as  
\begin{align}
    Y_{ij} = \mu + \gamma_{i} + \epsilon_{ij}, \label{eq:model1}
\end{align}
where $\mu$ is the grand mean rating, $\gamma_{i}$ is the ratee-specific deviation from the grand mean -- the  random intercept that together with the $\mu$ represents the ratees' true ratings -- with structural variance $\sigma^2_{\gamma}$, and finally, $\epsilon_{ij}$ is the random error of the rating with a residual variance $\sigma^2_{\epsilon}$. For the sake of estimation, it is standard to assume independent and identically distributed (IID) $\gamma_{i} \sim \textrm{N}(0, \sigma^2_{\gamma})$ and $\epsilon_{ij} \sim \textrm{N}(0, \sigma^2_{\epsilon})$, and random errors $\epsilon_{ij}$ to be uncorrelated with the ratee effects $\gamma_{i}$. Note that the assumption of IID error implies that there are different raters at each rating, or that the rater effect is neglected.

Under the model defined by Equation~\ref{eq:model1}, 
IRR is defined as the ratio of the true variance due to ratees, $\sigma^2_{\gamma}$, to the total variance $\sigma^2_{\gamma} + \sigma^2_{\epsilon}$, i.e., 
\begin{align}
    \textrm{IRR} = \frac{\sigma^2_{\gamma}}{\sigma^2_{\gamma} + \sigma^2_{\epsilon}},
    \label{eq:IRRmodel1}
\end{align}
corresponding to the intra-class correlation coefficient denoted as  ICC(1,1), see  \cite{shrout1979intraclass, mcgraw1996forming}, for more details and further possibilities.

The variance components in IRR defined by Equation~\ref{eq:IRRmodel1} can be estimated using various frequentist and Bayesian approaches to provide an estimate of IRR. The Maximum Likelihood (ML) estimates are found as parameters maximizing the likelihood function given the data \parencite{searle1997linear}. The REstricted (or REsidual) Maximum Likelihood (REML) method is an adaptation in which the marginal likelihood function is maximized with respect to variance components $(\sigma^2_\gamma, \sigma^2_\epsilon),$ while the third parameter $\mu$ is integrated out of the~likelihood function. For a balanced design of the model defined by Equation~\ref{eq:model1}, that is when the same number of ratings is given to all ratees, the REML estimates of variance components are identical to those from the one-way ANOVA method of moments \parencite{searle2006variance}. 

Finally, the Bayesian estimation \parencite[e.g., ][]{gelman2006data} starts with specifying prior distributions for all model parameters (the variance components $p(\sigma^2_\gamma, \sigma^2_\epsilon)$ and mean $p(\mu)$). The posterior distribution is then obtained through the Bayes' rule, by multiplying the prior distributions by the likelihood and standardizing by the probability of data (i.e., the marginal likelihood).

\subsubsection{Incorporating heterogeneous variance components}

The multilevel model defined by Equation~\ref{eq:model1} can be further generalized. The generalization suggested here involves the variance terms, possibly together with the mean, depending on covariates. For $Y_{ij}$ being the rating $j$ of subject $i$, we assume the following multilevel model
\begin{align}
    Y_{ij} = \mu_i + \gamma_{i} + \epsilon_{ij}, \label{eq:model}
\end{align}
where the mean $\mu_i,$ is modelled as a regression on covariates $\boldsymbol{x}_i$ 
\begin{align}
    \label{eq:parameterization}
    \mu_i                &= \alpha_{\mu} + \boldsymbol{\beta}_{\mu}^\top\boldsymbol{x}_i, 
\end{align}
$\gamma_i$ is a random effect of subject $i$, and $\epsilon_{ij}$ is a random error term for rating $j$ on subject $i$ as in Equation~\ref{eq:model1}. We moreover allow the covariates to influence the variance-terms. In other words, we assume that $\gamma_{i} \sim \textrm{N}(0, \sigma^2_{\gamma i})$ and $\epsilon_{ij} \sim \textrm{N}(0, \sigma^2_{\epsilon i})$ with  subject-specific variance terms $\sigma^2_{\gamma i},$ and $\sigma^2_{\epsilon i}$ being modelled as a regression on possibly a different set of covariates $\boldsymbol{u}_i$ and $\boldsymbol{v}_i$:
\begin{align}
    \label{eq:parameterization2}
    \sigma_{\epsilon i} &= \alpha_{\epsilon}  e^{\boldsymbol{\beta}_{\epsilon}^\top\boldsymbol{u}_i}, \\
    \nonumber
    \sigma_{\gamma i}   &= \alpha_{\gamma}  e^{\boldsymbol{\beta}_{\gamma}^\top\boldsymbol{v}_i}. 
\end{align}
In this equation, $\boldsymbol{\beta}_{\epsilon}$ are linear effects attributed to covariates $\boldsymbol{u}_i$ explaining the variability in variance terms $\sigma_{\epsilon i}$, while $\boldsymbol{\beta}_{\gamma}$ are linear effects attributed to covariates $\boldsymbol{v}_i$ explaining the variability in variance terms $\sigma_{\gamma i}$. The logarithmic link transforms the linear predictor into a multiplicative factor of the square root of the variance components' grand means, ensuring that the resulting variance is positive given that the grand means of the variance components are positive. 

Under a model  defined by Equation~\ref{eq:model}, the $\textrm{IRR}$ in Equation~\ref{eq:IRRmodel1} from Section~\ref{sec:meth:IRR} is generalized to depend on the set of covariates $\boldsymbol{u}_i$ and $\boldsymbol{v}_i$:

\begin{align}\label{eq:IRRgeneral}
    \textrm{IRR}_i = \textrm{IRR} (\boldsymbol{u}_i, \boldsymbol{v}_i) = \frac{\sigma^2_{\gamma i}}{\sigma^2_{\gamma i} + \sigma^2_{\epsilon i}} = \frac{\alpha_{\gamma}^2  e^{2\boldsymbol{\beta}_{\gamma}^\top\boldsymbol{v}_i}}{\alpha_{\gamma}^2  e^{2\boldsymbol{\beta}_{\gamma}^\top\boldsymbol{v}_i} + \alpha_{\epsilon}^2  e^{2\boldsymbol{\beta}_{\epsilon}^\top\boldsymbol{u}_i}}. 
\end{align}

Note that when including covariates $\boldsymbol{x}_i$ in the fixed part of the model defined in Equation~\ref{eq:parameterization}, the estimates as well as the meaning of the variances in Equations~\ref{eq:parameterization2} change, and the interpretation of IRR in Equation~\ref{eq:IRRgeneral} changes as well. More specifically, by including overall effects of ratee characteristics in a model, the between-ratee variance will typically be reduced in some groups and the group differences in $\sigma_\gamma^2$ will become smaller. As a practical consequence, the interpretation of the IRR in such a case relates to an instance whereby the final judgment takes into consideration the covariates. As an example, \cite{goldhaber2021profreferences} assumed that hiring officials were likely to take the rater type into consideration when interpreting the professional references’ ratings of teacher applicants, and thus their inter-rater reliability estimates were adjusted for these sources of variation by including the rater type into the fixed part of the model. If, on the contrary, the final judgments are completed based solely upon the ratings, a simpler version of the model given by Equation~\ref{eq:model} should be considered, in which $\mu$ is a constant, i.e., $\boldsymbol{\beta}_{\mu}$ in Equation~\ref{eq:parameterization} is restricted to 0.

Also note that we generally allow a possibly different set of covariates $\boldsymbol{x}_i$, $\boldsymbol{u}_i$, and $\boldsymbol{v}_i$ to explain the means $\mu_i$, structural variances $\sigma_{\gamma i}$, and residual variances $\sigma_{\epsilon i}$ in Equation~\ref{eq:model}, that we will use in the real data example in Section~\ref{sec:RealDataNIH2}. However, in certain situations, a simpler and more restrictive model may be assumed, in which the same set of covariates is used, as we will do in our simulation study in Section~\ref{sec:Simulation} and in the real data examples in Section~\ref{sec:RealDataAIBS} and~\ref{sec:RealDataNIH1}.

\subsection{Bayesian hypothesis testing and model-averaging}

With Equation~\ref{eq:model} denoting the most complex model, a number of submodels can be considered as special cases based on restricting some of the effects in $\boldsymbol{\beta}_\mu, \boldsymbol{\beta}_\gamma, \boldsymbol{\beta}_\epsilon$ to zero. With a number of models to select from, we can select the best fitting model (as discussed in this subsection) and use parameter estimates for the best-fitting model for the final estimate of IRR. Alternatively, we can incorporate the uncertainty of the model selection process and calculate the model-averaged parameter estimates.

\subsubsection{Bayes factors}
\label{sec:ModelSelectionBF}

We consider here the Bayesian hypothesis testing framework of \textcite{jeffreys1931scientific} which evaluates the evidence in support of / against any model by the usage of Bayes factors. Bayes factors are computed as a ratio of the marginal likelihoods of the competing models \parencite{wrinch1921on, kass1995bayes, etz2017haldane, rouder2019teaching}
\begin{equation}
  \label{eq:BF_simple}
  \text{BF}_{10} = \frac{p(\text{data} \given \mathcal{M}_{1})}{p(\text{data} \given \mathcal{M}_{0})},
\end{equation}
\noindent with the marginal likelihood $p(\text{data} \given \mathcal{M}_{m})$ quantifying the model's $m$ relative predictive performance by integrating the likelihood over the parameter space \parencite{jefferys1992ockhams}. 
%
%

The Bayes factor is a continuous measure of evidence in favor of $\mathcal{M}_{1}$ and against $\mathcal{M}_{0}$. For ease of interpretation, we can label the resulting Bayes factors as weak ($\text{BF}_{10}$ between 1 and 3), moderate (between 3 and 10), strong (between 10 and 100), and very strong (larger than 100) (\cite[Appendix I]{Jeffreys1939}; \cite{kass1995bayes}; \cite[p. 105]{LeeWagenmakersBayesBook}).

\subsubsection{Bayesian model-averaging}
\label{sec:BMA}

In addition to model selection and using the single best fitting model for parameter estimation, we should also consider Bayesian model-averaging \parencite{leamer1978specification, hoeting1999bayesian, kass1995bayes}.

Bayesian model-averaging accounts for the uncertainty of model selection by weighting the posterior model estimates by posterior model probabilities. First, we need to assign prior model probabilities $p(\mathcal{M}_{m})$ to the individual models $m$ and update them with the Bayes' rule into posterior model probability $p(\mathcal{M}_{m} \given \text{data})$ according to the Bayes' rule \parencite{hoeting1999bayesian, fragoso2018bayesian, hinne2019conceptual},
\begin{equation}
     p(\mathcal{M}_{m} \given \text{data}) = \frac{p(\text{data} \given \mathcal{M}_{m}) \times p(\mathcal{M}_{m})}{\sum_{m = 1}^M p(\text{data} \given \mathcal{M}_{m}) \times p(\mathcal{M}_{m})}.
\end{equation}

We then combine the posterior parameter estimates $p(\theta \given \text{data}, \mathcal{M}_m)$ from the $m = 1, \dots, M$ individual models based on posterior model probabilities $p(\mathcal{M}_m \given \text{data})$,
\begin{equation}
    p(\theta \given \text{data}) = \sum_{m = 1}^{M} p(\theta \given \mathcal{M}_{m}, \text{data}) \times p(\mathcal{M}_{m} \given \text{data}),
\end{equation}
\noindent which allows us to acknowledge the uncertainty about the considered models. We follow a common convention in Bayesian model-averaging and assign an equal prior model probability to models assuming the absence and presence of the difference between the groups for either the mean, structural, or residual variance, resulting in $p(\mathcal{M}_{m}) = \nicefrac{1}{M}$ \parencite{kass1995bayes, gronau2021primer, madigan1994strategies, raftery1995accounting}.

Furthermore, Bayesian model-averaging allows us to quantify evidence in favor of including a specific parameter across the whole set of specified models with a comparable structure. E.g., for a difference between the groups in a residual variance $\sigma^2_\epsilon$, the Bayes factor from Equation~\ref{eq:BF_simple} is extended into \emph{inclusion Bayes factor} \parencite{gronau2021primer, hinne2019conceptual}, 
\begin{equation}
    \label{eq:inclusion_BF}
    \small
    \underbrace{ \text{BF}_{\epsilon \overline{\epsilon}} }_{ \substack{\text{Inclusion Bayes factor}\\{\text{for difference in } \epsilon}} } =  
    \;\;\;  \underbrace{ \frac{ \sum_{a \in A} p(\mathcal{M}_{a} \given \text{data}) }
    { \sum_{b \in B} p(\mathcal{M}_{b} \given \text{data}) }}_{ \substack{\text{Posterior inclusion odds}\\{\text{for difference in }\epsilon}}} \;\;\; \Bigg/ \underbrace{ \frac{ \sum_{a \in A} p(\mathcal{M}_{a}) }
    { \sum_{b \in B} p(\mathcal{M}_{b}) }}_{\substack{\text{Prior inclusion odds}\\{\text{for difference in }\epsilon}}},
\end{equation}
where $A$ represents a set of models for which the groups differ in the $\sigma^2_\epsilon$ parameter, and $B$ represents a set of models for which they don't differ.

\subsubsection{Parametrization and choice of priors}

To employ the Bayesian framework consisting of Bayes factors and Bayesian model-averaging, we need to complete the models by specifying prior distributions for all the parameters in Equations~\ref{eq:parameterization} and ~\ref{eq:parameterization2}. Here, we restrict ourselves to consideration of binary covariates, testing for and quantifying the differences between groups (see Discussion for suggestions on dealing with other types of covariates). We use effect coding (i.e., we assign the values of -0.5 and 0.5 for the two levels), so the prior distribution on the regression coefficients $\beta$ corresponds to the difference (for the mean rating) or standard deviation ratio (for the structural and residual variances) between the groups. Consequently, the intercept parameters $\alpha$ represent the unweighted grand means, i.e., they have common interpretation across all possible submodels.

For the simulation study and the real-data example, we use the following priors
\begin{align}
    \label{eq:stan-parameterization}
    \alpha_{\mu}  &\sim \text{Normal}(0, \, 1),  \\
    \nonumber
    \alpha_{\gamma}, \, \alpha_{\epsilon}  &\sim \text{Normal}_+(0, \, 1),  \\
    \nonumber
    \beta_{\mu}, \, \beta_{\gamma}, \, \beta_{\epsilon}  &\sim \text{Normal}(0, \, 0.5^2),
\end{align}

where $\text{Normal}_+$ stands for the half-normal distribution, and 
\begin{align*}
    \gamma_i             &\sim \text{Normal}(0, \, \sigma_{\gamma i}^2), \\
    y_{ij}               &\sim \text{Normal}(\mu_i + \gamma_i, \, \sigma_{\epsilon i}^2),
\end{align*}
where $\mu_i$ is defined by Equation~\ref{eq:parameterization} with $\boldsymbol{x} = -0.5$ for the first group, and $\boldsymbol{x} = 0.5$ for the second group, and $\sigma_{\gamma i}^2$ and $\sigma_{\epsilon i}^2$ are defined by Equation~\ref{eq:parameterization2} for each participant $i$.

Our reasoning behind the choice of priors is as follows: Since the intercept parameters are common across all models (i.e., we are not going to test for the presence or absence of the intercept), we can specify weakly informative prior distributions on them \parencite{gelman2006data}. Here, we use standard normal prior distribution for the grand mean intercept, $\alpha_\mu \sim N(0,1)$, and half normal prior distributions for the structural and residual standard deviation intercepts, $\alpha_\gamma, \alpha_\epsilon \sim N_+(0,1)$. This setting corresponds to the expectation that the outcome variable is somewhat standardized, i.e., the grand mean is located around zero and the overall variance of the data is around one. If the outcome variable corresponded to a differently scaled measure, we would adjust the means and standard deviations of the prior distributions to reflect the overall expectations (e.g., we could use $\alpha_\mu \sim N(100, 15^2)$ and $\alpha_\gamma, \alpha_\epsilon \sim N_+(0, 15^2)$ if we were working with IQ scores).

In contrast to the common intercepts, the regression parameters $\beta$ can differ between the submodels (omitting a predictor equals to setting the corresponding $\beta = 0$). Subsequently, the prior distribution on the regression parameters defines the hypothesis test for the presence or the absence of the effect for a given predictor. Here, we use informed Normal$(0, \sigma^2)$ prior distributions on the regression coefficients where $\sigma^2$ parameter controls informativeness (i.e., deviations from the null hypotheses we are interested in) of the test. This corresponds to specifying a two sided hypothesis on the regression parameters for the means and standard deviations. In our view, the choice of $\sigma^2 = 0.5^2$ used in the simulation study and the real data example corresponds to testing for ``medium sized'' differences in means and standard deviation ratios (i.e., mean differences lower than 1, and standard deviation ratios lower than 2.7). 

\begin{figure}[h!]
    \centering
    \includegraphics[width=\textwidth]{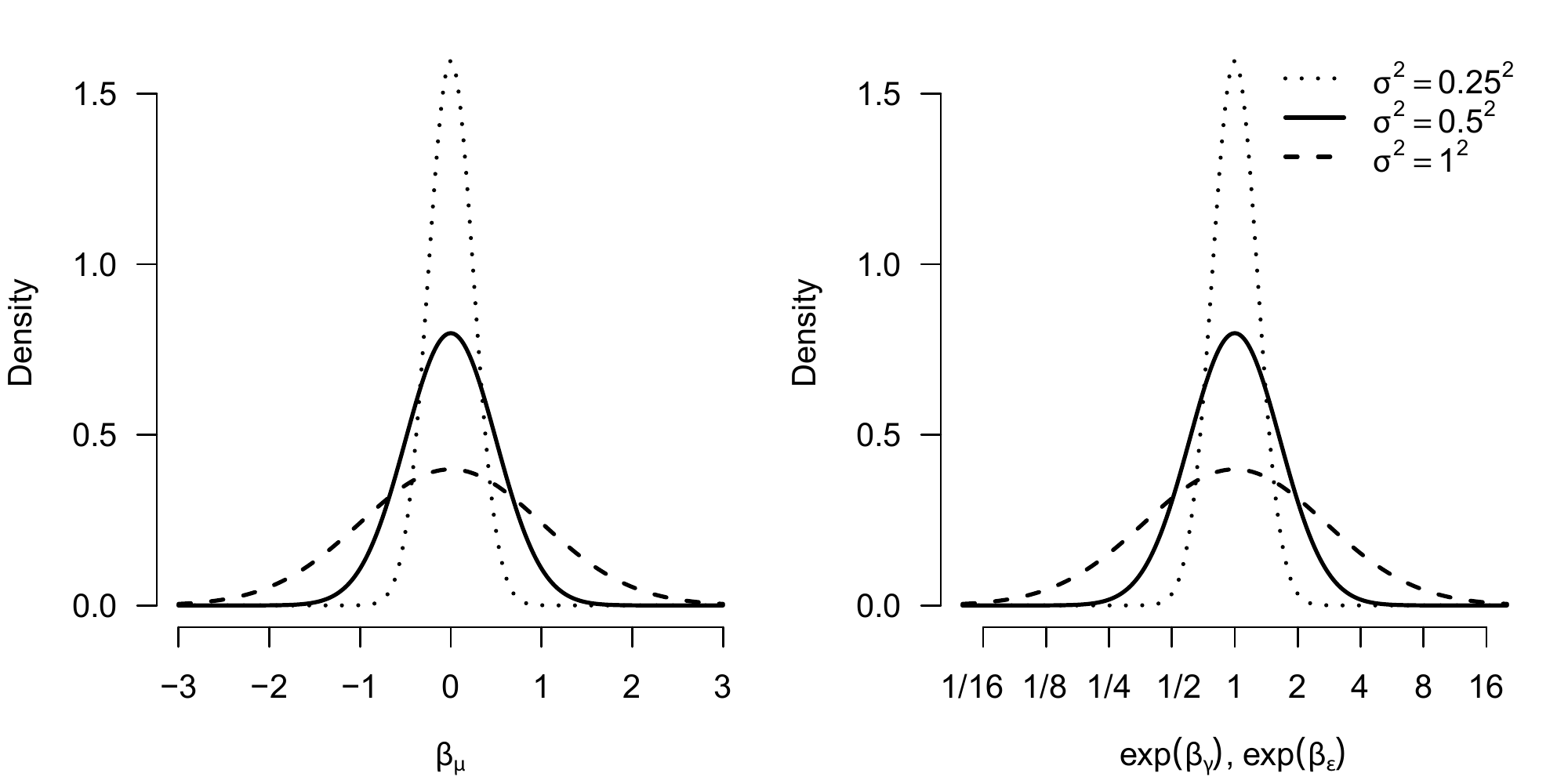}
    \caption{Visualization of different options of prior distributions on the regression coefficients. We used the Normal(0, $0.5^2$) prior distributions (in bold) for the simulation and the real data example and the remaining options as a robustness check. The left panel visualizes the resulting prior distribution for the mean differences and the right panel for standard deviation ratios. }\label{fig:prior_distributions}
\end{figure}

To assess the robustness of our results to the prior distribution specifications, we use two other choices of $\sigma^2$ in the real data example. In our view, the choices of $\sigma^2 = 0.25^2$ and $\sigma^2 = 1^2$ correspond to ``small sized'' and ``large sized'' differences in means and standard deviation ratios, respectively. See Figure~\ref{fig:prior_distributions} for the considered resulting prior distributions of the mean differences (left panel) and the resulting prior distributions of the ratios of standard deviations (right panel) obtainable by taking the exponent of the prior distribution. 

\section{Simulation study}
\label{sec:Simulation}

We perform a simulation study to assess performance of the outlined methodology. We are specifically interested in an estimation and hypothesis testing in consideration to the differences between groups in any of the modeled parameters (i.e., means, structural standard deviations, and residual standard deviations) and IRR. We keep simulation settings simple, in order to compare the outlined methodology to other model selections and model-averaging techniques, considering that many of the other methods could not deal with more complex data settings.

For the simulation, we consider IRR in a group-specific variance components model, in other words, we assume a single binary covariate, the group membership denoted by index $g \in \{0, 1\}$. 
For $Y_{ijg}$ being the rating $j$ of subject $i$ from group $g$, the original model defined by Equation~\ref{eq:model}  simplifies to
\begin{align}
    Y_{ijg} = \mu_g + \gamma_{ig} + \epsilon_{ijg}, \label{eq:modelG}
\end{align}
where $\mu_g$ is the group-specific mean rating, $\gamma_{ig} \sim \textrm{N}(0, \sigma^2_{\gamma g})$ is ratee-specific deviation from the group mean with a group-specific (structural) variance $\sigma^2_{\gamma g}$, and finally, $\epsilon_{ijg} \sim \textrm{N}(0, \sigma^2_{\epsilon g})$ is the random error of the rating with group-specific residual variance $\sigma^2_{\epsilon g}$. We assume normal distributions.

Under the model specified by Equation \ref{eq:modelG}, the group-specific $\textrm{IRR}_g$ is then defined in the special case of Equation \ref{eq:IRRgeneral} as \begin{align}
    \textrm{IRR}_g = \frac{\sigma^2_{\gamma g}}{\sigma^2_{\gamma g} + \sigma^2_{\epsilon g}} 
\end{align}
and it takes the two values of $\textrm{IRR}_0$ and $\textrm{IRR}_1$ depending upon the group, which is the only covariate assumed in this design.

Possible submodels of the model defined by Equation~\ref{eq:modelG} are derived as special cases based upon restricting the group-specific parameters under a combination of conditions 
\begin{align*}
    \textrm{A. }&\mu_0 = \mu_1 = \mu, \\ 
    \textrm{B. }& 
    \sigma^2_{\gamma 0} = \sigma^2_{\gamma 1} = \sigma^2_{\gamma}, \\ 
    \textrm{ C. }& 
    \sigma^2_{\epsilon 0} = \sigma^2_{\epsilon 1} = \sigma^2_{\epsilon}. 
\end{align*}
Altogether, the combination of conditions A, B, and C leads to 7 possible submodels denoted as M1 -- M7, and 1 unrestricted model (Equation~\ref{eq:modelG}) denoted as M8.
\begin{align*}
    \textrm{M1: } & [\textrm{A} \textrm{B} \textrm{C}] & \quad 
    & Y_{ijg} = \mu + \gamma_{i} + \epsilon_{ij}\\
    \textrm{M2: } & [\textrm{A} \textrm{B} \bar{\textrm{C}}] & \quad & Y_{ijg} = \mu + \gamma_{i} + \epsilon_{ijg}\\
    \textrm{M3: } & [\textrm{A} \bar{\textrm{B}} \textrm{C}] & \quad 
    & Y_{ijg} = \mu + \gamma_{ig} + \epsilon_{ij}\\
    \textrm{M4: } & [\textrm{A} \bar{\textrm{B}} \bar{\textrm{C}}] & \quad & Y_{ijg} = \mu + \gamma_{ig} + \epsilon_{ijg}\\
    \textrm{M5: } & [\bar{\textrm{A}} \textrm{B} \textrm{C}] & \quad 
    & Y_{ijg} = \mu_g + \gamma_{i} + \epsilon_{ij}\\
    \textrm{M6: } & [\bar{\textrm{A}} \textrm{B} \bar{\textrm{C}}] & \quad & Y_{ijg} = \mu_g + \gamma_{i} + \epsilon_{ijg}\\
    \textrm{M7: } & [\bar{\textrm{A}} \bar{\textrm{B}} \textrm{C}] & \quad 
    & Y_{ijg} = \mu_g + \gamma_{ig} + \epsilon_{ij}\\
    \textrm{M8: } & [\bar{\textrm{A}} \bar{\textrm{B}} \bar{\textrm{C}}] & \quad & Y_{ijg} = \mu_g + \gamma_{ig} + \epsilon_{ijg}
\end{align*}

\subsection{Data generation} 
Data generation was inspired by real data encountered in the context of teacher applicant ratings \parencite{martinkova2018disparities}. 
Specifically, in Equation~\ref{eq:modelG}, we used two values for the standardized mean differences between the groups ($\mu_2 - \mu_1 = 0, \textrm{ or } 0.4$), for the structural variance ratios $\left( \nicefrac{\sigma^2_{\gamma 1}}{\sigma^2_{\gamma 2}} = 1, \textrm{ or } 1.5\right)$, and for residual variance ratios $\left( \nicefrac{\sigma^2_{\epsilon 1}}{\sigma^2_{\epsilon 2}} = 1, \textrm{ or } 1.5\right)$, while we constrained the overall mean variance across groups to  $\nicefrac{1}{G} \sum_{g=1}^G \sigma_{\gamma g}^2 + \sigma_{\epsilon g}^2 = 1$, and the mean IRR across groups to $\nicefrac{1}{G} \sum_{g=1}^G \textrm{IRR}_g = 0.45$. This led to eight simulation scenarios, with scenarios 4 and 8 split into two sub-scenarios depending upon whether the structural and residual variance ratios differed in the same or the opposite direction (Table~\ref{tab:SimulationSettings}). 

\begin{table}[h!]
\centering
\small
\caption{Data Generation Scenarios}
\begin{tabular}{lrrrrrrrr}
  \hline
  Scenario & $\mu_1$ & $\mu_2$ & $\sigma_{\gamma 1}$ & $\sigma_{\gamma 2}$ & $\sigma_{\epsilon 1}$ & $\sigma_{\epsilon 2}$ & $\text{IRR}_{1}$ & $\text{IRR}_{2}$ \\ 
  \hline
  1        & 0.00 & 0.00 & 0.67 & 0.67 & 0.74 & 0.74 & 0.45 & 0.45 \\
  2        & 0.00 & 0.00 & 0.67 & 0.67 & 0.67 & 0.82 & 0.50 & 0.40 \\
  3        & 0.00 & 0.00 & 0.60 & 0.74 & 0.74 & 0.74 & 0.40 & 0.50 \\
  4.1      & 0.00 & 0.00 & 0.60 & 0.73 & 0.66 & 0.81 & 0.45 & 0.45 \\
  4.2      & 0.00 & 0.00 & 0.73 & 0.60 & 0.66 & 0.81 & 0.55 & 0.35 \\
  5        &-0.20 & 0.20 & 0.67 & 0.67 & 0.74 & 0.74 & 0.45 & 0.45 \\
  6        &-0.20 & 0.20 & 0.67 & 0.67 & 0.67 & 0.82 & 0.50 & 0.40 \\
  7        &-0.20 & 0.20 & 0.60 & 0.74 & 0.74 & 0.74 & 0.40 & 0.50 \\
  8.1      &-0.20 & 0.20 & 0.60 & 0.73 & 0.66 & 0.81 & 0.45 & 0.45 \\
  8.2      &-0.20 & 0.20 & 0.73 & 0.60 & 0.66 & 0.81 & 0.55 & 0.35 \\
  \hline
\end{tabular}\label{tab:SimulationSettings}
\end{table}

Moreover, we manipulated the number of times 
the ratees were rated (J = 3, \textrm{ or } 5) and the number of ratees per group (I = 25, 50, 100, \textrm{ or } 200) in each scenario. In total, 10 (scenarios including subscenarios) $\times$ 2 (number of ratings) $\times$ 4 (number of ratees) = 80 conditions were simulated, 1000 times each, implying 80,000 randomly generated data sets.

\subsection{Compared methods}

We compared the Bayesian hypothesis testing and model-averaging methodology outlined in the Methods sections to alternative frequentist and Bayesian ways of estimating and testing for the differences in group-specific mixed-effects location scale models defined by Equation~\ref{eq:modelG}. Specifically, we used the maximum likelihood (ML) and restricted maximum-likelihood (REML) estimation in the frequentist framework for linear mixed models and Markov chain Monte Carlo (MCMC) estimation in the Bayesian framework. 

\paragraph{Model selection. }
To assess the performance of the model selection with Bayes factors specified in Section~\ref{sec:ModelSelectionBF}, we considered four frequentist model selection approaches and two Bayesian approaches.

The frequentist approaches include two stepwise selection procedures (backward and forward) and two model space selection procedures based upon Akaike information criterion \parencite[AIC,][]{akaike1974new} and the Bayesian information criterion \parencite[BIC,][]{schwarz1978estimating}.
In the forward stepwise selection procedure, we started with the simplest model. We first tested for adding the difference in means (with REML) and then gradually tested expanding the model with differences in structural and/or residual variances (with ML; see the left panel of Figure~\ref{fig:model_selection} in Appendix~A for diagram). 
In the backward stepwise selection procedure, we started with the most complex model. We first tested for gradually removing differences in the structural and/or residual variances (with ML) and then tested for removing the difference in means (with REML; see the right panel of Figure~\ref{fig:model_selection} in Appendix~A for the diagram). 
In the model selection procedures based upon information criteria, we estimated all the specified models (with ML) and subsequently selected the best fitting model based on the lowest information criteria (AIC or BIC).

The Bayesian approaches include two model space selection procedures based upon posterior predictive performance --- the Watanabe–Akaike Information Criterion \parencite[WAIC,][]{watanabe2010asymptotic} and Leave-One-Out cross-validation \parencite[LOO,][]{VehtariEtAl2017}. WAIC and LOO approximate the leave-one-out prediction error using the log-likelihood evaluated with posterior parameter distribution \parencite{mcelreath2018statistical}. While they are asymptotically equivalent, LOO usually perform better in small samples and under weak prior distributions \parencite{VehtariEtAl2017}. In our case, we use the individual ratings $Y_{ij}$ as the basis of leave-one-out predictions.\footnote{We found that leave-one-rating cross-validation performed better than leave-one-ratee cross-validation in our simulations, therefore, we show results only for leave-one-rating based metrics.}

\paragraph{Model-averaging. }
To assess the performance of the parameter estimation with Bayesian model averaging, we used two frequentist and two alternative Bayesian approaches. 

The frequentist methods combine the estimates $\hat{\theta}_m$ from $M$ individual models with weights $\omega_m$,
\begin{equation}
  \hat{\theta} = \sum_{m = 1}^{M} \hat{\theta}_m \times \omega_m.
\end{equation}
\noindent 
The two frequentist approaches are based on information criteria (AIC, BIC), and specify the weight $\omega_m$ for model $m$ as 
\begin{align}
  \label{eq:AIC_weights}
  \omega_{\text{AIC}, m} &= \frac{ \text{exp}(-\nicefrac{1}{2} \, \Delta_m(\text{AIC}) ) }{ \sum_{i = 1}^{M} \text{exp}(-\nicefrac{1}{2} \, \Delta_i(\text{AIC}) ) }, \\
  \nonumber
  \omega_{\text{BIC}, m} &= \frac{ \text{exp}(-\nicefrac{1}{2} \, \Delta_m(\text{BIC}) ) }{ \sum_{i = 1}^{M} \text{exp}(-\nicefrac{1}{2} \, \Delta_i(\text{BIC}) ) }, 
\end{align}
with
\begin{align*}
    \Delta_m(\text{AIC}) = \text{AIC}_m - \text{min}(\text{AIC}),\\
    \Delta_m(\text{BIC}) = \text{BIC}_m - \text{min}(\text{BIC}),
\end{align*}
where $\text{AIC}_m$ and $\text{BIC}_m$ correspond to the AIC and BIC value of the $m^\text{th}$ model and $\Delta_m(\text{AIC}_m)$ and $\Delta_m(\text{BIC}_m)$ correspond to the difference between the AIC or BIC, respectively, of the $m^\text{th}$ and the best fitting model \parencite{hjort2003frequentist, wagenmakers2004aic}.

As an alternative Bayesian approach, the pseudo Bayesian model-averaging, similarly to the frequentist model-averaging, uses information criteria to compute the model weights in Equation~\ref{eq:AIC_weights} \parencite{geisser1979predictive, gelfand1996model}. In contrast to Bayesian model-averaging, it does not require specification of prior model probabilities, since the weights are based entirely on the LOO information criteria.

The last alternative approach, the Bayesian stacking of predictive distributions \parencite{yao2018using} is based upon stacking which combines models in order to minimise leave-one-out mean square error \parencite{wolpert1992stacked, leblanc1996combining, breiman1996stacked}. The stacking of posterior distribution is then based upon the leave-one-out predictive distribution computed with LOO. Similarly to pseudo-Bayesian model-averaging, Bayesian stacking does not require specification of prior model probabilities, however, it oftentimes does not allow inference about the true data structure and is unable to provide compelling evidence in favor of simple models \parencite{GronauWagenmakers2019LOO1, GronauWagenmakers2019LOO2}.

\subsubsection{Evaluation of the simulation results}
We first evaluate the proportion of selecting the correct model based upon Bayes factors and we compare it with other approaches. We evaluate the proportion of correct model selection averaged across all conditions and separately for each of the data generating model and sample size.

Next, we compare our approach with other approaches in the precision of estimates of model parameters and of IRR. As a measure of precision, we evaluate the root mean square error (RMSE). As a (relative) measure of bias, we also evaluate the bias$^2$/MSE ratio. This is again evaluated when averaged across all conditions, as well as for all the individual model generating designs.

Finally, we evaluate the calibration of our prior distributions and of inclusion Bayes factors with the 
so-called Bayes factor design analysis \parencite[see, e.g.,][
for more details]{stefan2019tutorial, schonbrodt2018bayes}. Our goal for this simulation
was (1) to verify that the inclusion Bayes factors found evidence in favor of the difference in parameters in those conditions where the parameters differed and evidence in favor of no difference in conditions where the parameters do not differ, (2) to evaluate the proportion of misleading evidence, i.e., how often would the inclusion Bayes factors find strong evidence in favor of a difference in scenarios with no difference present, and finally (3) to verify that the evidence is increasing with an increasing sample size.

\subsection{Implementation}
The simulation was carried out in \texttt{R} version 3.5.1 \parencite{R}. We used \texttt{nlme} \texttt{R} package version 3.1 \parencite{nlme} to estimate the frequentist version of the models and we have written a custom Rstan model with the usage of \texttt{rstan} \texttt{R} package version 2.18.2 \parencite{Rstan} for the Bayesian models. We further used the \texttt{bridgesampling} \texttt{R} package version 3.1 \parencite{GronauEtAl2020JSS} to compute the marginal likelihoods via bridge sampling \parencite[e.g.,][]{GronauEtAl2017BSTutorial, MengWong1996}, and we used \texttt{loo} \texttt{R} package version 2.0 \parencite{loo} to compute WAIC, LOO, and pseudo-BMA and stacking weights with the usage of Pareto-smoothed importance sampling \parencite{VehtariEtAl2017}.

\subsection{Simulation results}
\label{sec:Results}

\subsubsection{Model selection}
We first take a look at the averaged results across all conditions. The probability of selecting the correct model is summarized in Table~\ref{tab:ProbCOrrectModel3}. Bayes factors, AIC, WAIC, and LOO are able to identify the correct model with precision around 26\% of the time in the smallest group sizes (N = 25), however, while Bayes factors and AIC steadily improve with increasing sample sizes (up to 70\% and 65\% respectively), WAIC and LOO start lagging behind. Furthermore, the step-wise forward and backward selection procedures start catching up with the Bayes factors with increasing sample sizes (69\%). 

\begin{table}[h!]
\centering
\caption{Proportion (and Standard Error of Proportion) of the Correctly Selected Models (Averaged across Conditions, Number of Ratings per Rated Subject $j$ = 3).}
\begin{tabular}{lcccc}
  \hline
  Method     & N = 25        & N = 50        & N = 100       & N = 200       \\ 
  \hline
  BF         & 0.259 (0.005) & 0.379 (0.005) & 0.545 (0.006) & 0.704 (0.005) \\ 
  AIC        & 0.270 (0.005) & 0.399 (0.005) & 0.537 (0.006) & 0.652 (0.005) \\ 
  BIC        & 0.199 (0.004) & 0.283 (0.005) & 0.424 (0.006) & 0.589 (0.006) \\ 
  forward    & 0.223 (0.005) & 0.349 (0.005) & 0.526 (0.006) & 0.687 (0.005) \\ 
  backward   & 0.223 (0.005) & 0.350 (0.005) & 0.528 (0.006) & 0.687 (0.005) \\ 
  WAIC       & 0.264 (0.005) & 0.372 (0.005) & 0.479 (0.006) & 0.552 (0.006) \\ 
  LOO        & 0.262 (0.005) & 0.374 (0.005) & 0.480 (0.006) & 0.560 (0.006) \\
  \hline
\end{tabular}\label{tab:ProbCOrrectModel3}
\end{table}

\begin{figure}[h!]
    \centering
    \includegraphics[width=\textwidth]{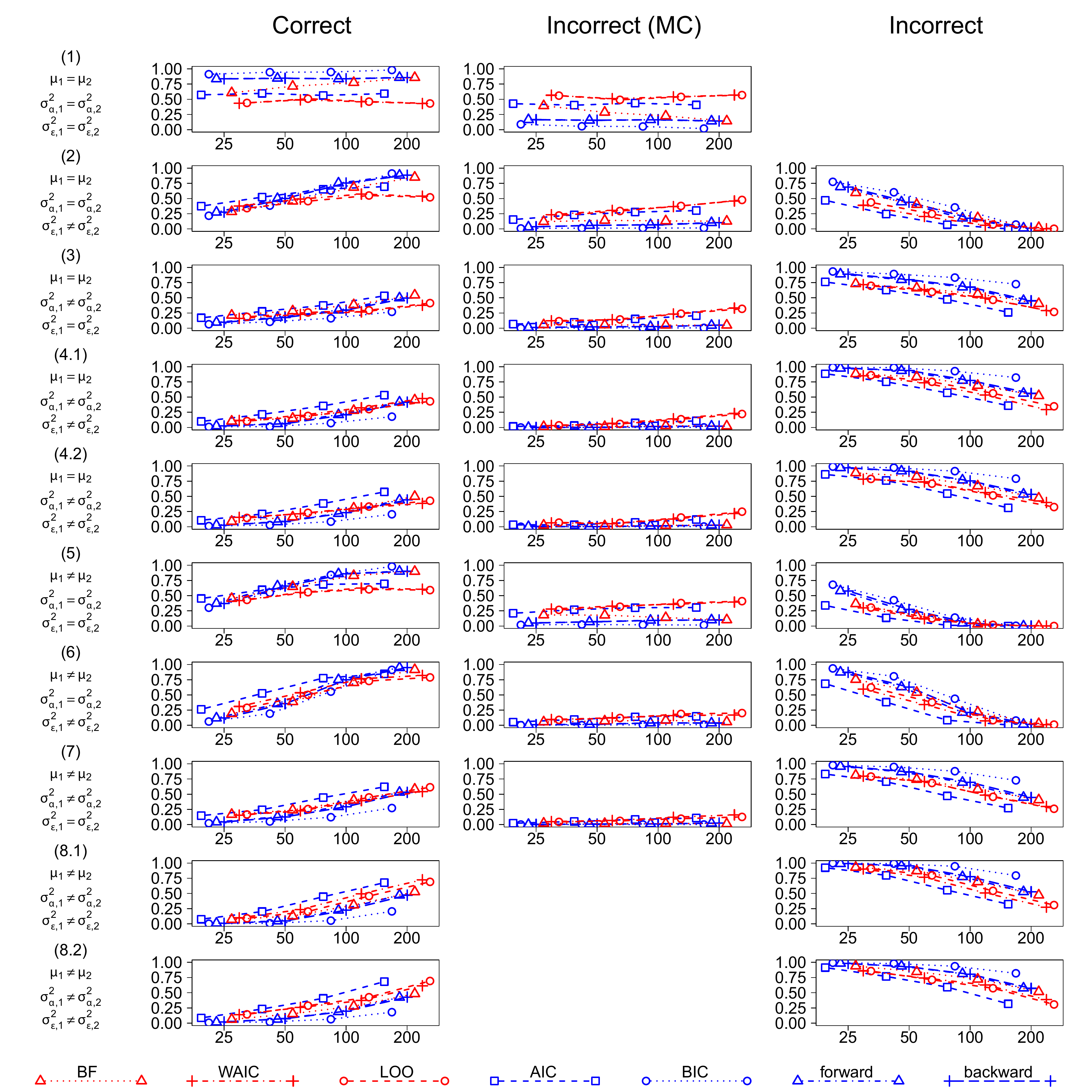}
    \caption{Proportion of correct, incorrect - more complex (MC, containing all true parameter differences plus some additional incorrect differences) and incorrect (missing at least one parameter difference) model selection. Red: Bayesian, Blue: Frequentist model selection techniques. BF: Model selection with Bayes factor proposed here.}\label{fig:model_selection_n3}
\end{figure}

Figure~\ref{fig:model_selection_n3} displays model selection performance for individual data generating designs.  The first column shows the proportion of correctly selected models, the second column shows the proportion of selecting a more complex incorrect model (i.e., a model containing all true parameter differences plus some additional incorrect differences), and the last column shows the proportion of other incorrect models (models missing at least one parameter difference). We see a well-known behavior of BIC being biased towards the simpler model, resulting in ``better'' performance under data-generating scenario 1, and a bias of WAIC and LOO towards more complex models, resulting in a lower proportion of correct model selection and increased selection of incorrect more complex models in the simpler data generating designs. The same trends are visible in the case of $j = 5$ ratings per ratee, see the electronic Supplementary Material.

\subsubsection{Parameter estimation} 

The RMSE of the residual SD estimates, averaged across all conditions, is summarized in  Table~\ref{tab:RMSEresid3}.  We can see that with small samples (e.g., N = 25), the model averaging leads to more precise estimates of residual variance than the estimates based upon the selected best-performing model. For example, Bayesian parameter estimation in the models selected with Bayes factors resulted in RMSE of 0.075 while the Bayesian model averaging resulted in RMSE of 0.069. With a growing sample size, the uncertainty in model selection disappears, and the RMSE of the two approaches converge to the same value. The same trend can be seen for the IRR estimates in Table~\ref{tab:RMSEirr3}, however, the benefits of model averaging are less pronounced than in the case of residual variances. 

\begin{table}[h!]
\centering
\caption{RMSE (and Standard Error of RMSE) of the Residual SD Estimates (Averaged across Conditions, Number of Ratings per Rated Subject $j$ = 3).}
\begin{tabular}{lcccc}
  \hline
            & N = 25     & N = 50     & N = 100    & N = 200    \\ 
  \hline
  \multicolumn{5}{l}{Model Selection}                           \\
  BF        & 0.075 (0.000) & 0.056 (0.000) & 0.039 (0.000) & 0.025 (0.000) \\ 
  AIC       & 0.075 (0.000) & 0.053 (0.000) & 0.037 (0.000) & 0.025 (0.000) \\ 
  BIC       & 0.076 (0.000) & 0.060 (0.000) & 0.044 (0.000) & 0.026 (0.000) \\ 
  forward   & 0.076 (0.000) & 0.057 (0.000) & 0.039 (0.000) & 0.025 (0.000) \\ 
  backward  & 0.076 (0.000) & 0.057 (0.000) & 0.039 (0.000) & 0.025 (0.000) \\ 
  WAIC      & 0.073 (0.000) & 0.053 (0.000) & 0.037 (0.000) & 0.025 (0.000) \\ 
  LOO       & 0.074 (0.000) & 0.053 (0.000) & 0.037 (0.000) & 0.025 (0.000) \\ 
  Full model& 0.074 (0.000) & 0.052 (0.000) & 0.037 (0.000) & 0.026 (0.000) \\ 
  \hline
  \multicolumn{5}{l}{Model Averaging}                           \\
  BMA       & 0.069 (0.000) & 0.052 (0.000) & 0.038 (0.000) & 0.025 (0.000) \\ 
  AIC       & 0.069 (0.000) & 0.051 (0.000) & 0.037 (0.000) & 0.025 (0.000) \\ 
  BIC       & 0.070 (0.000) & 0.054 (0.000) & 0.041 (0.000) & 0.026 (0.000) \\ 
  WAIC      & 0.070 (0.000) & 0.051 (0.000) & 0.036 (0.000) & 0.025 (0.000) \\ 
  pseudoBMA & 0.069 (0.000) & 0.051 (0.000) & 0.037 (0.000) & 0.025 (0.000) \\ 
  stacking  & 0.071 (0.000) & 0.051 (0.000) & 0.036 (0.000) & 0.025 (0.000) \\ 
   \hline
\end{tabular} \label{tab:RMSEresid3}
\end{table}

\begin{table}[h!]
\centering
\caption{RMSE (and Standard Error of RMSE) of the IRR Estimates (Averaged across Conditions, Number of Ratings per Rated Subject $j$ = 3).}
\begin{tabular}{rllll}
  \hline
            & N = 25     & N = 50     & N = 100    & N = 200    \\ 
  \hline
  \multicolumn{5}{l}{Model Selection}                           \\
  BF        & 0.106 (0.001) & 0.082 (0.000) & 0.060 (0.000) & 0.044 (0.000) \\
  AIC       & 0.119 (0.001) & 0.086 (0.000) & 0.061 (0.000) & 0.043 (0.000) \\ 
  BIC       & 0.110 (0.001) & 0.084 (0.000) & 0.064 (0.000) & 0.048 (0.000) \\ 
  forward   & 0.112 (0.001) & 0.085 (0.000) & 0.062 (0.000) & 0.045 (0.000) \\ 
  backward  & 0.112 (0.001) & 0.085 (0.000) & 0.062 (0.000) & 0.045 (0.000) \\ 
  WAIC      & 0.106 (0.001) & 0.081 (0.000) & 0.059 (0.000) & 0.043 (0.000) \\ 
  LOO       & 0.106 (0.001) & 0.081 (0.000) & 0.059 (0.000) & 0.043 (0.000) \\
  \hline
  \multicolumn{5}{l}{Model Averaging}                           \\
  BMA       & 0.100 (0.001) & 0.076 (0.000) & 0.056 (0.000) & 0.042 (0.000) \\ 
  AIC       & 0.108 (0.001) & 0.080 (0.000) & 0.057 (0.000) & 0.041 (0.000) \\ 
  BIC       & 0.102 (0.001) & 0.078 (0.000) & 0.059 (0.000) & 0.045 (0.000) \\ 
  WAIC      & 0.100 (0.001) & 0.076 (0.000) & 0.055 (0.000) & 0.041 (0.000) \\ 
  pseudoBMA & 0.099 (0.001) & 0.075 (0.000) & 0.055 (0.000) & 0.041 (0.000) \\ 
  stacking  & 0.103 (0.001) & 0.077 (0.000) & 0.056 (0.000) & 0.041 (0.000) \\ 
  Full model& 0.122 (0.001) & 0.087 (0.000) & 0.061 (0.000) & 0.043 (0.000) \\ 
   \hline
\end{tabular} \label{tab:RMSEirr3}
\end{table}

Analogous tables with the RMSE of the mean estimates and structural variances, which are not the main focus of IRR and our study, are available in the electronic Supplementary Material, as well as the corresponding results for $j = 5$ ratings.
The electronic Supplementary Material also provides figures with more detailed results for individual data generating models. Namely, for the case of $j = 3$ and $j = 5$ ratings per ratee, we depict the bias, RMSE, and ratio of bias$^2$/MSE for estimates of the means, structural and residual variances, as well as IRR under different data-generating models. We can see a fairly similar performance in the terms of bias and RMSE across methods with a decreasing bias and RMSE with sample size and higher benefits of model-averaging in smaller samples. The bias$^2$/MSE then illustrates the bias variance trade-off between model-averaging and model selection, where the decrease in RMSE is accompanied with a relative increase in bias.

\subsubsection{Inclusion Bayes factor calibration}
Finally, we turn our attention to performance of the inclusion Bayes factors in providing evidence for differences in the mean, the structural and the residual variance between the two groups.
Table~\ref{tab:BF_calibration_3} summarizes the proportion of inclusion Bayes factors correctly favoring the true data generating model for each parameter in both types of scenarios (with difference vs. without difference in a given parameter), averaged across data-generating scenarios of a given type,
and the rate of misleading strong evidence ($\text{BF}_{10} > 10$ in the case of no difference or $\text{BF}_{10} < \nicefrac{1}{10}$ in the case of a difference between the groups), in brackets. The table suggests that the inclusion Bayes factors are well calibrated for providing evidence about the differences in means but are noticeably biased when providing evidence towards no difference in structural variances, and slightly biased towards models with no difference in residual variances in small samples. However, the rate of misleading strong evidence is minimal, with less than 0.6\% across the parameters and conditions. The proportion of Bayes factors correctly favoring the true data generating process is quickly increasing with the sample size, reaching the probability of 95\% for differences in means and for differences in residual variances with $n = 200$. The case of $j = 5$ ratings depicted in the electronic Supplementary Material then shows that the bias towards no difference in a case of differences in structural variances is improving with the increased number of ratings, pointing to a lack of information about the structural variances themselves in cases of a low number of ratings, which favors the simpler models (i.e., models assuming no difference in structural variance).
  
\begin{table}[h!]
\centering
\caption{Proportion (and Standard Error of Proportion) of Inclusion Bayes Factors Favoring the Correct Data-Generating Mechanism and the Rate of Misleading Strong Evidence (in Brackets); Averaged across Different Data-Generating Scenarios, Number of Ratings per Rated Subject $j$ = 3.}
\begin{tabular}{lcccc}
  \hline
  Scenario     & N = 25        & N = 50        & N = 100       & N = 200        \\ 
  \hline
  \multicolumn{5}{l}{Mean ($\mu$)}                                              \\ 
  Difference    & 0.626 (0.000) & 0.814 (0.000) & 0.961 (0.000) & 1.000 (0.000) \\ 
  No difference & 0.846 (0.006) & 0.888 (0.006) & 0.923 (0.005) & 0.944 (0.003) \\ 
  \hline
  \multicolumn{5}{l}{Structural variances ($\sigma_{\gamma}$)}                  \\ 
  Difference    & 0.277 (0.000) & 0.316 (0.000) & 0.402 (0.000) & 0.543 (0.000) \\ 
  No difference & 0.805 (0.004) & 0.858 (0.006) & 0.894 (0.006) & 0.933 (0.004) \\ 
  \hline
  \multicolumn{5}{l}{Residual variances ($\sigma_{\epsilon}$)}                  \\
  Difference    & 0.399 (0.000) & 0.586 (0.000) & 0.809 (0.000) & 0.971 (0.000) \\ 
  No difference & 0.907 (0.004) & 0.936 (0.004) & 0.952 (0.002) & 0.968 (0.003) \\ 
  \hline
\end{tabular}\label{tab:BF_calibration_3}
\end{table}

\begin{figure}[h!]
    \centering
    \includegraphics[width=\textwidth]{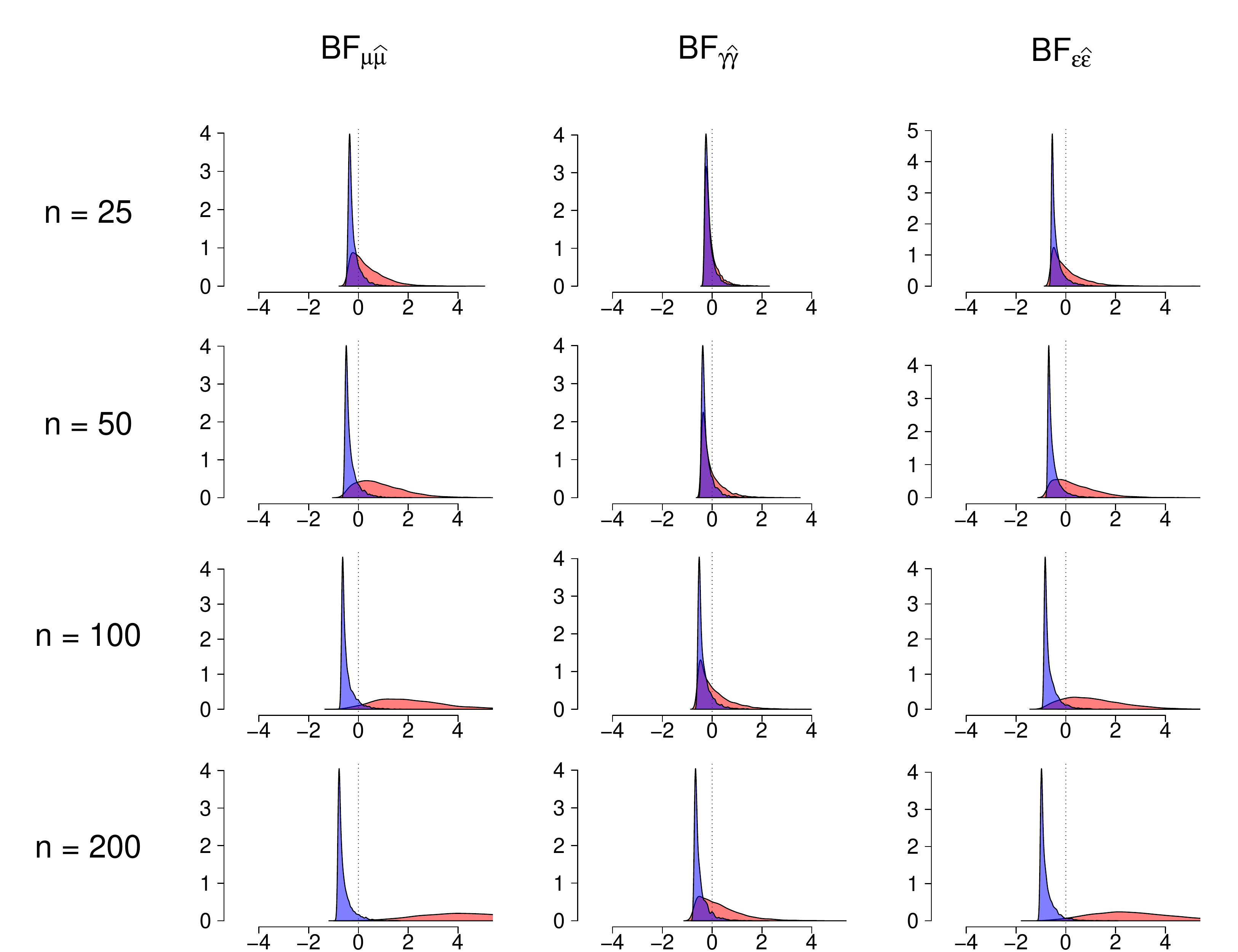}
    \caption{Visualization of $\text{log}_{10}$ of the inclusion Bayes factors for each parameter aggregated across  sample size $n$. Positive values of $\text{log}_{10}$ inclusion Bayes factors correspond to evidence in favor of difference in the parameter, negative values correspond to evidence in favor of no difference in the parameter. Red: aggregated inclusion Bayes factors for scenarios with \textbf{difference} for a given parameter. Blue: distribution visualizing the aggregated inclusion Bayes factors for scenarios with \textbf{no difference} for a given parameter conditions.}\label{fig:BF_calibration_3}
\end{figure}

A more detailed behavior of the inclusion Bayes factors is depicted in Figure~\ref{fig:BF_calibration_3}, visualizing the distribution of ($\text{log}_{10}$) inclusion Bayes factors for a difference in each parameter (columns) with increasing sample sizes (rows) across all simulation conditions. Inclusion Bayes factors from data-generating scenarios with a difference in a given parameter are depicted in red and inclusion Bayes factors from scenarios with no difference in a given parameter are depicted in blue. 

We can see that inclusion Bayes factors for differences in means $\text{BF}_{\mu, \bar{\mu}}$ and residual variances $\text{BF}_{\epsilon, \bar{\epsilon}}$ quickly converge at the correct solution, i.e. to the right side of the figure in case of a difference (depicted in red), and to the left side of the figure in case of no difference (depicted in blue). 
However, the inclusion Bayes factors for the differences in structural variances $\text{BF}_{\gamma, \bar{\gamma}}$ converge at the correct solution much slower due to the considerably smaller amount of information about structural variances provided by only 3 ratings per subject. See the electronic Supplementary Material for a comparison with the case of 5 rating per subject that shows a sightly improved performance.

\section{Real data examples}\label{sec:RealDataExample}
We demonstrate the estimation of IRR with two datasets of grant proposal peer-review analyzed by \textcite{erosheva2021irr} and available in the \texttt{ShinyItemAnalysis} \texttt{R} package \parencite{martinkova2018shinyitemanalysis}. 

Unlike in the simulation presented in Section~\ref{sec:Simulation}, with the real data, the true generating model is unknown. However, the practical examples provide an illustration regarding the application of the proposed method; for the first two examples incorporating a single covariate, we can also compare the results with those provided by other methods presented in the simulation study.

In our analysis, we assume that the selection panels may take the applicant's gender and career stage into consideration when interpreting the ratings, i.e., we allow covariates to explain the fixed part $\mu_i$ in Equation~\ref{eq:model}). In Appendix~C and in the Supplementary Material, we include the results if no covariate adjustment is expected when interpreting the ratings by the selection panel (i.e., when only models with a constant $\mu$ are considered).

\subsection{AIBS grant proposal review data with single covariate}
\label{sec:RealDataAIBS}

The first example involves peer-review ratings of the American Institutes of Biological Sciences (AIBS), analyzed by \textcite{erosheva2021irr}. In the AIBS dataset, each grant proposal was rated three times on the number of criteria as well as on the overall merit score considered here as the dependent variable. The gender (n\textsubscript{female} = 25, n\textsubscript{male} = 47)  of the principal investigator was used as a covariate / grouping variable.

The Bayes factor, as well as other model selection methods (both Bayesian and frequentist), indicated the simplest model (M1) was the most suitable for the data (Table~\ref{tab:AIBS}). However, there was a relatively large uncertainty about the selected model, as can be seen from Table~\ref{tab:weights_AIBS} which summarizes the weights for the individual models. The simulation study suggested that with this small sample size, all methods have a low probability of selecting the correct model, and it is hard to judge which of the models is true. The simulation also demonstrated that the model selection based upon BIC more often prefers simpler models, which can also be observed in our practical example, where the model weight based upon BIC is much higher (0.78) for model M1 than the weights for this model based upon other criteria.

\begin{table}[h!]
\small
\centering
\caption{AIBS Peer-Review Example: IRR Estimates for Female (G1) and Male (G2) Principal Investigators.}
\begin{tabular}{lrrr|rrr|rrr}
    \hline
            & $\text{IRR}_{G1}$ & LCI & UCI & $\text{IRR}_{G2}$ & LCI & UCI & $\triangle$ICC & LCI & UCI  \\ 
    \hline
    \multicolumn{10}{l}{Model Selection} \\
    BF/WAIC/LOO (M1)       & 0.37 & 0.22 & 0.52 & 0.37 & 0.22 & 0.52 & 0 & 0 & 0 \\ 
    AIC/BIC/forw/back (M1) & 0.37 & 0.22 & 0.51 & 0.37 & 0.22 & 0.51 & 0 & 0 & 0 \\ 
    \hline
    \multicolumn{10}{l}{Model-Averaging} \\
    BMA       & 0.40 & 0.22 & 0.60 & 0.35 & 0.18 & 0.51 & 0.05 & -0.08 & 0.30 \\ 
    AIC       & 0.40 & 0.20 & 0.60 & 0.34 & 0.18 & 0.51 & 0.06 & -0.16 & 0.27 \\ 
    BIC       & 0.37 & 0.21 & 0.54 & 0.36 & 0.21 & 0.51 & 0.01 & -0.10 & 0.12 \\ 
    WAIC      & 0.40 & 0.22 & 0.61 & 0.35 & 0.18 & 0.51 & 0.05 & -0.08 & 0.31 \\ 
    stacking  & 0.37 & 0.22 & 0.52 & 0.37 & 0.22 & 0.52 & 0.00 &  0.00 & 0.00 \\ 
    pseudoBMA & 0.40 & 0.22 & 0.60 & 0.35 & 0.18 & 0.51 & 0.05 & -0.09 & 0.30 \\ 
   \hline
\end{tabular}\label{tab:AIBS}
\end{table}

\begin{table}[ht]
\centering
\caption{AIBS Peer-Review Example: Model Weights. \\ \emph{Note:} Stacking weights are not displayed due to their high variability and dependence on a simulation run (with the majority of the stacking weights assigned to Model 1 and Model 3 across 5 different MCMC initialization conditions).}

\begin{tabular}{lrrrrrrrr}
  \hline
  Method    & M1   & M2   & M3   & M4   & M5   & M6   & M7   & M8   \\ 
  \hline 
  BMA       & 0.34 & 0.09 & 0.24 & 0.07 & 0.12 & 0.03 & 0.09 & 0.03 \\ 
  AIC       & 0.32 & 0.13 & 0.18 & 0.09 & 0.12 & 0.05 & 0.07 & 0.03 \\ 
  BIC       & 0.78 & 0.06 & 0.08 & 0.01 & 0.05 & 0.00 & 0.01 & 0.00 \\ 
  WAIC      & 0.21 & 0.12 & 0.16 & 0.09 & 0.15 & 0.08 & 0.11 & 0.06 \\ 
  pseudoBMA & 0.24 & 0.11 & 0.17 & 0.09 & 0.16 & 0.07 & 0.11 & 0.06 \\ 
  \hline
\end{tabular}\label{tab:weights_AIBS}
\end{table}

Quantifying the evidence across all models with inclusion Bayes factors resulted only in weak evidence in support of the absence of difference in the means, $\text{BF}_{\overline{\mu} \mu} = 2.74$, weak evidence in support of the absence of difference in structural variances $\text{BF}_{\overline{\gamma} \gamma} = 1.38$, and moderate evidence in support of the absence of difference in  residual variances, $\text{BF}_{\overline{\epsilon} \epsilon} = 3.52$. In other words, there was no clear evidence supporting differences in the means or in structural variances between the two gender groups, but the data were more consistent with differences in residual variances between the two gender groups.

\subsection{NIH  grant proposal review data with single covariate}
\label{sec:RealDataNIH1}
For the second example, we used the  data of peer-review ratings of the National Institutes of Health (NIH), analyzed by \cite{erosheva2020nih} and further discussed in \cite{erosheva2021irr}. We used the preliminary Investigator criterion scores, and the gender (n\textsubscript{female} = 574, n\textsubscript{male} = 1310) of the principal investigator as a covariate; 

All model selection methods selected models which assumed a difference in residual variances: Bayes factors and BIC suggested model M2, the model selection based upon LOO, WAIC, forward, and backward selection also suggested a difference in means (model M6), while the method based upon AIC selected the most complicated model also suggesting a difference in structural variances. According to models M2 and M6, the single-rater IRR for grant proposals by male PIs is 0.36 (0.33 -- 0.39), whereas for grant proposals by female PIs it is 0.33 (0.29/0.30 -- 0.36) with a difference in IRR of 0.04 (0.01/0.02 -- 0.06). However, according to model M8, the difference of IRRs for male and female PIs is insignificant -0.01 and with a wider CI of (-0.07 -- 0.05) also covering zero, see the top panel of Table~\ref{tab:IRR_NIH_sex}.

Despite  a much larger sample size than in the previous example, there is still considerable uncertainly about the selected model (see Table~\ref{tab:MAweightsNIHsex}). We again use model averaging to account for the model uncertainty, which results in wider confidence intervals. Most of the approaches find differences in IRR between the two groups insignificant, with a confidence interval of $\triangle$IRR covering zero, see the last three columns of the bottom part of Table~\ref{tab:IRR_NIH_sex}.

\begin{table}[h!]
\centering
\caption{NIH Peer-Review IRR Estimates Investigator Scale of the Complete Data Set of ``Male'' (G1) and ``Female'' (G2) Principal Investigators. IRR estimates for G1, G2, and the difference of IRR ($\triangle$IRR) are complemented with lower (LCI) and upper (UCI) bounds of 95\% confidence intervals for the frequentist models and of 95\% central credible intervals for the Bayesian models.}
\begin{tabular}{rrrr|rrr|rrr}
    \hline
            & $\text{IRR}_{G1}$ & LCI & UCI & $\text{IRR}_{G2}$ & LCI & UCI & $\triangle$IRR & LCI & UCI  \\ 
    \hline
    \multicolumn{10}{l}{Model Selection} \\

    BF (M2)            & 0.36 & 0.33 & 0.39 & 0.33 & 0.30 & 0.36 & 0.04 & 0.02 & 0.06 \\ 
    AIC (M8)           & 0.35 & 0.31 & 0.38 & 0.36 & 0.31 & 0.41 & -0.01 & -0.07 & 0.05 \\ 
    BIC (M2)           & 0.36 & 0.33 & 0.39 & 0.33 & 0.30 & 0.36 & 0.04 & 0.02 & 0.06 \\ 
    forw/back (M6)     & 0.36 & 0.33 & 0.39 & 0.33 & 0.29 & 0.36 & 0.04 & 0.01 & 0.06 \\ 
    LOO/WAIC (M6)      & 0.36 & 0.33 & 0.39 & 0.33 & 0.29 & 0.36 & 0.04 & 0.01 & 0.06 \\
    \hline
    \multicolumn{10}{l}{Model-Averaging} \\ 
    BMA                & 0.36 & 0.32 & 0.39 & 0.34 & 0.30 & 0.41 & 0.01 & -0.08 & 0.06 \\ 
    AIC                & 0.35 & 0.32 & 0.39 & 0.35 & 0.29 & 0.41 & 0.00 & -0.07 & 0.08 \\ 
    BIC                & 0.36 & 0.32 & 0.39 & 0.34 & 0.29 & 0.39 & 0.02 & -0.04 & 0.08 \\ 
    WAIC               & 0.36 & 0.32 & 0.39 & 0.34 & 0.30 & 0.40 & 0.01 & -0.07 & 0.06 \\  
    stacking           & 0.35 & 0.31 & 0.39 & 0.35 & 0.30 & 0.43 & 0.00 & -0.10 & 0.06 \\ 
    pseudoBMA          & 0.35 & 0.31 & 0.39 & 0.35 & 0.30 & 0.42 & 0.00 & -0.09 & 0.05 \\ 
    \hline
\end{tabular}\label{tab:IRR_NIH_sex}
\end{table}

\begin{table}[ht]
\centering
\caption{Model-Averaging Weights for NIH Peer-Review IRR Estimates Investigator Scale of the Complete Data Set of ``Male'' (G1) and ``Female'' (G2) Principal Investigators.}
\begin{tabular}{lrrrrrrrr}
  \hline
  Method    & M1 & M2 & M3 & M4 & M5 & M6 & M7 & M8 \\ 
  \hline 
  BMA       & 0.01 & 0.36 & 0.05 & 0.11 & 0.02 & 0.30 & 0.04 & 0.11 \\ 
  AIC       & 0.00 & 0.07 & 0.01 & 0.11 & 0.00 & 0.32 & 0.03 & 0.46 \\ 
  BIC       & 0.17 & 0.60 & 0.06 & 0.03 & 0.03 & 0.10 & 0.01 & 0.01 \\ 
  WAIC      & 0.01 & 0.24 & 0.00 & 0.16 & 0.01 & 0.32 & 0.01 & 0.24 \\ 
  pseudoBMA & 0.09 & 0.17 & 0.07 & 0.10 & 0.05 & 0.29 & 0.07 & 0.14 \\ 
  \hline
\end{tabular}\label{tab:MAweightsNIHsex}
\end{table}

Quantifying the evidence across all models with inclusion Bayes factors resulted only in weak evidence in support of the absence of difference in the means, $\text{BF}_{\overline{\mu} \mu} = 1.15$, weak evidence in support of the absence of difference in structural variances $\text{BF}_{\overline{\gamma} \gamma} = 2.21$, and moderate evidence in support of the presence of difference in residual variances, $\text{BF}_{\epsilon \overline{\epsilon}} = 7.25$. In other words, there was again no clear evidence in favor of differences in the means or in structural variances between the two gender groups, but the data were more consistent with differences in residual variances between the two gender groups. Acknowledging that the differences in residual variances may be caused by differences in career stage, we investigate further in the next section.

\subsection{NIH  grant proposal review data with more covariates}\label{sec:RealDataNIH2}

We finally considered a more complex situation of IRR being dependent upon two covariates: gender and binarized career stage. In this case, the model given by Equation~\ref{eq:modelG} has 64 sub-models. Model selection using Bayes factors identified that the data were best predicted by a model in which the means, structural variances, and residual variances differed by career stage, with the posterior model probability of 0.23. Table~\ref{tab:NIH_best_models} shows ten models which were best at predicting the data, accumulating a total of 95\% of the posterior model probabilities. We can see that despite the large sample size and the considerable uncertainty about the best model, most of the best performing models consider the career stage to be an important predictor for all parameters. However, gender does not seem to play a central role and is slightly more often not even considered.

\begin{table}[ht]
\centering
\caption{Model structure for the ten best performing models in the NIH data set when considering both the gender and career stage as predictors. 
\emph{Note:} First three columns describe the model in terms of predictors of each parameter ($\mu$, $\sigma_\gamma$, and $\sigma_\epsilon$). Marg. Lik denotes the marginal likelihood, $p(\mathcal{M}_{i})$ the prior model probability, and $p(\mathcal{M}_{i} \given \text{data})$ the posterior model probability of each model.}
\label{tab:NIH_best_models}
\begin{tabular}{cccrrr}
  \hline
  $\mu$ & $\sigma_\gamma$ & $\sigma_\epsilon$ & Marg. Lik. & $p(\mathcal{M}_{i})$ & $p(\mathcal{M}_{i} \given \text{data})$ \\ 
  \hline
  Stage           & Stage           & Stage           & -7672.06 & 0.02 & 0.23 \\ 
  Stage           & Gender \& Stage & Stage           & -7672.15 & 0.02 & 0.21 \\ 
  Stage           & Stage           & Gender \& Stage & -7672.28 & 0.02 & 0.19 \\ 
  Stage           & Gender \& Stage & Gender \& Stage & -7673.39 & 0.02 & 0.06 \\ 
  Stage           & Gender          & Stage           & -7673.47 & 0.02 & 0.06 \\ 
  Gender \& Stage & Stage           & Stage           & -7673.77 & 0.02 & 0.04 \\ 
  Stage           & None            & Stage           & -7673.79 & 0.02 & 0.04 \\ 
  Gender \& Stage & Gender \& Stage & Stage           & -7673.87 & 0.02 & 0.04 \\ 
  Stage           & None            & Gender \& Stage & -7673.98 & 0.02 & 0.03 \\ 
  Gender \& Stage & Stage           & Gender \& Stage & -7673.99 & 0.02 & 0.03 \\ 
  \hline
\end{tabular}
\end{table}

An overall picture is provided using Bayesian model-averaging which combines estimates and evidence across all models. We find weak evidence against difference in residual variance between the two gender groups, $\text{BF}_{\epsilon, \bar{\epsilon}} = \nicefrac{1}{1.82} = 0.55$, however, we find very strong evidence for the difference between the two career stage groups in residual variance, $\text{BF}_{\epsilon, \bar{\epsilon}} = 1.09 \times 10^{17}$. The model-averaged estimates of the residual standard deviation ratios are 0.98 (95\% central credible interval: 0.91 -- 1.00) for the two gender groups and 0.78 (0.74 -- 0.83) for the two career stage groups.

While the \emph{residual variance} is a parameter of central importance for the assessment of measurement error and inter-rater reliability, we may also derive from the results conclusions regarding \emph{structural variance}, the \emph{mean}, and regarding between-group differences in these parameters. We find weak evidence against difference in structural variance between the two gender groups, $\text{BF}_{\bar{\gamma}, \gamma} = \nicefrac{1}{1.43} = 0.70$, and we find moderate evidence for difference between the two career stage groups, $\text{BF}_{\gamma, \bar{\gamma}} = 4.74$. The model-averaged posterior mean estimates of the structural standard deviation ratios are 0.96 (0.81 -- 1.00) for the two gender groups and 0.85 (0.72 -- 1.00) for the two career stage groups. We find moderate evidence in favor of no difference in means by gender $\text{BF}_{\bar{\mu}, \mu} = \nicefrac{1}{0.18} = 5.57$ and very strong evidence for difference in mean by career stage $\text{BF}_{\mu, \bar{\mu}} = 1.24 \times 10^{31}$. The model-averaged posterior mean estimates of the differences in mean are $-$0.01 ($-$0.08 -- 0.00) for the two gender groups, and $-$0.56 ($-$0.65 -- $-$0.48) for the two career stage groups.

Higher residual variance in the nonexperienced group is accompanied by only slightly higher structural variance, and it leads to a somewhat lower IRR, see Table~\ref{tab:NIH_parameters}. Note that when only the models with no covariate effect on the mean are considered, the structural variance in the nonexperienced group is higher, leading to higher IRR in this group, see Table~\ref{tab:NIH_parameters_unadjusted} in Appendix~C.

\begin{table}[ht]
\centering
\caption{Estimated Model-Averaged Marginal Means and 95\% CI for Each of the Parameters. \\ \emph{Note:} Exp -- Experienced, nExp -- Non-Experienced}
\begin{tabular}{rrrrrr}
  \hline
  Gender & Stage & \multicolumn{1}{c}{$\mu$} & \multicolumn{1}{c}{$\sigma_\gamma$} & \multicolumn{1}{c}{$\sigma_\epsilon$} & \multicolumn{1}{c}{IRR} \\ 
  \hline
  Female & nExp & 0.44 [0.37, 0.51] & 0.63 [0.52, 0.73] & 0.98 [0.93, 1.04] & 0.29 [0.21, 0.37] \\ 
  Male & nExp & 0.43 [0.37, 0.49] & 0.60 [0.50, 0.68] & 0.96 [0.91, 1.00] & 0.28 [0.20, 0.34] \\ 
  Female & Exp & -0.12 [-0.19, -0.05] & 0.53 [0.47, 0.61] & 0.76 [0.73, 0.81] & 0.33 [0.27, 0.40] \\ 
  Male & Exp & -0.13 [-0.20, -0.07] & 0.51 [0.45, 0.56] & 0.75 [0.72, 0.78] & 0.32 [0.26, 0.37] \\ 
   \hline
\end{tabular}\label{tab:NIH_parameters}
\end{table}

We further conducted a sensitivity analysis to assess how our conclusions would change if we specified different prior distributions, i.e., tested different hypotheses. We used the two remaining prior distributions depicted in Figure~\ref{fig:prior_distributions}; (1) a more concentrated prior distribution around no effect with the standard deviation $\sigma = 0.25$ -- testing for the presence of smaller differences between the groups and (2) a wider prior distribution with standard deviation $\sigma = 1$ -- testing for the presence of larger differences between the groups.

We will only discuss here the residual variances, see Appendix~B for more details. Between the two gender groups, we found strong evidence for the absence of larger differences in the residual variances and no evidence supporting the presence or the absence of small differences. For the groups segregated by career stage, we found very strong evidence for presence of the effect, regardless of the specification of the alternative hypothesis. In other words, the data were more consistent with no or small differences in residual variances between the two gender groups and there was clear evidence in favor of differences in residual variances between the two career stage groups.

\section{Discussion}\label{sec:Discussion}

In this work, we have presented a new flexible approach for assessing the IRR in cases where variance components, depending upon covariates, are assumed to differ. We used mixed effect models with heterogeneous variances and employed the Bayesian framework with Bayes factors and Bayesian model-averaging. In a simulation study, we compared the methodology to other frequentist and Bayesian approaches and shown comparable or superior performace to those of the other methods.  More importantly, flexibility in the proposed methodology allows researchers to straightforwardly extend the presented models to cases with more covariates. Whereas Bayes factors can be used to select a single model, researchers can further account uncertainty in the model structure with Bayesian model-averaging when drawing inferences about either the presence or absence of the effect via inclusion.  

The suggested methodology -- Bayesian hypothesis testing and model-averaging -- is, of course, not the only option researchers can pursue. Authors with different philosophical views would advocate for different approaches, such as estimation only or inference based upon confidence intervals \parencite[e.g.,][]{gelman2006data, cumming2014new}. We prefer the Bayesian hypothesis testing since we believe that Bayes factors (and likelihood ratio tests) are the only coherent method of testing for the presence vs. absence of the effect. The problem with hypothesis testing based upon posterior credible intervals (or $p$-values) is the assumption of either the presence (for credible intervals) or absence (for $p$-values) of the effect at the onset of the analysis. In other words, it is impossible to provide evidence for/against an assumption that is already taken for granted \parencite[e.g.,][]{Jeffreys1939}.

The advantages of Bayesian hypothesis testing and model-averaging however come at an additional cost: the specification of prior distributions. Prior distributions are especially important upon the parameters of interest where they define the hypotheses about the presence vs. the absence of an effect. Different prior distributions equal to different hypotheses -- different questions -- being asked. Subsequently, different questions might lead to different answers (e.g., a one-sided vs. two-sided test). Nonetheless, as shown by the sensitivity analysis in Appendix~B, similar prior distributions correspond to similar questions which subsequently result in similar answers. To define a prior distribution, researchers must be able to define the degree of effects they are interested in, see e.g., \textcite{johnson2010methods, ohagan2006uncertain, mikkola2021prior} for detailed information about prior distribution elicitation. 

The proposed method was further demonstrated when assessing IRR in a grant proposal peer-review with respect to the applicant's gender and career stage. The results suggested that the IRR is not likely dependent upon the gender of the principal investigator, while it may be lower with a lower career stage. When demonstrated in this specific example of grant peer review, it is worth noting the importance and the wide range of possible applications using the proposed method. Our methods may be used to identify gaps in the IRR for various rating situations (applicant hiring or promotion, classroom observation of teachers, journal peer-review, etc.) and with respect to the different types of rater and ratee characteristics (dichotomous such as internal/external status, factors such as social status or marital status, and continuous such as age). 

We also discussed how hypotheses regarding specific variance components may be addressed with the inclusion Bayes factors. This is especially important because IRR is influenced by range restriction \parencite{erosheva2021irr}, meaning that for a fixed residual variance, different values of IRR are obtained depending upon the structural variance. For this reason, the difference between residual variances may be of greater interest than the difference in the IRR itself. This aspect was also demonstrated with the NIH example by comparing the case of when the grand mean was allowed to vary with covariates to the case of no covariate adjustment: In the latter case, the structural variance for the Non-Experienced group was higher, leading to a somewhat higher IRR than in the Experienced group, while, when the ratings accounted for the stage, the structural variance was somewhat lower for the Non-Experienced group, leading to a somewhat lower IRR. Unlike IRR which provided contradictory conclusions, the inclusion Bayes factors provided more coherent information and in both cases unanimously concluded there is a significant difference in the residual variance and more error in the ratings from applicants in the Non-Experienced group.

Several limitations of the current study and possible directions for future research are worth mentioning: First, a simulation study will always cover only a finite and rather limited number of parameter setups, and our simulation study involved only one binary covariate. Nevertheless, the current simulation study was already extensive in terms of the number of methods compared and the simulation time needed. Besides the computation time, some aspects of the frequentist approaches would need to be solved. As an example, the parametrization is not straightforward in the \texttt{lme()} and \texttt{lmer()} functions for cases involving additional covariates of variance components. The model selection outlined in Appendix~A also becomes more complicated with the addition of covariates of variance components and the actual size of the entire space for possible models increases exponentially. Our real data examples show how a model with one covariate can be applied to models with multiple covariates.

Secondly, we considered binary covariates only. However, our approach could be easily extended to other types of covariates. For example, in the case of factors with multiple groups,  one has to decide whether an ANOVA-like test for at least one difference between the factor levels should be specified with orthonormal contrasts asserting that the prior marginal levels are identical, making the levels interchangable \parencite{rouder2012default}, or, whether a multiple treatments vs control-like test should be specified with dummy coding, asserting that the control condition corresponds to the grand mean and the coefficients for each treatment condition to differences / standard deviation ratios. Similarly, 
priors on continuous covariates simply correspond to the unit change in the covariate, with the grand mean / variance corresponding to the covariate value of 0, while the centering or re-scaling of the covariate prior to the analysis might simplify the prior specification.

Thirdly, we considered only the simplest model with the ratee being the single structural source of error, while all other possible sources of error (such as rater, occasion, etc.) were encompassed in the residual error. This model is appropriate and widely used when most of the raters rate only a single ratee. In real-life applications, the hierarchy may be more sophisticated (respondents nested within institutions, which themselves are nested within towns or countries), and there may be more sources of error such as raters, so-called facets in the context of the Generalizability theory \parencite{Brennan2001}. Different generalizability and dependability coefficients may then be defined in such cases, and the IRR of interest may need to be defined by more complex ratios with different interpretations. However, the cases of heterogeneity would then be treated analogously, and the Bayesian approach suggested here would be easily applied to more complex situations.

Regardless of the limitations, the study offers a flexible method for assessing the heterogeneity in IRR with respect to rater and ratee characteristics. This may help identifying the subgroups with lower IRR and improving the IRR of these groups and in general. This in turn may be of great importance to those designing the ratings and to policy makers whose interest is to improve assessment systems and the selection processes.

\section*{Acknowledgement}

The study was funded by the Czech Science Foundation grant number 21-03658S. We appreciate the computing and storage facilities of the Institute of Computer Science (Czech Republic RVO 67985807) and access to computing and storage facilities owned by parties and projects contributing to the National Grid Infrastructure MetaCentrum provided under the program “Projects of Large Research, Development, and Innovations Infrastructures” (CESNET LM2015042). We are thankful to Martin Otava, Jon Kern, and anonymous reviewers for helpful comments and suggestions on earlier versions of the manuscript. 
\section*{Supplementary Material}

Additional tables and figures, and accompanying \texttt{R} scripts are available at \href{https://osf.io/bk8a7/}{https://osf.io/bk8a7/}.


\printbibliography

\newpage
\appendix
\beginappendix


\section*{Appendix A: Model selection diagrams}

\begin{figure}[h!]
    \begin{center}
        \includegraphics[width=0.40\textwidth]{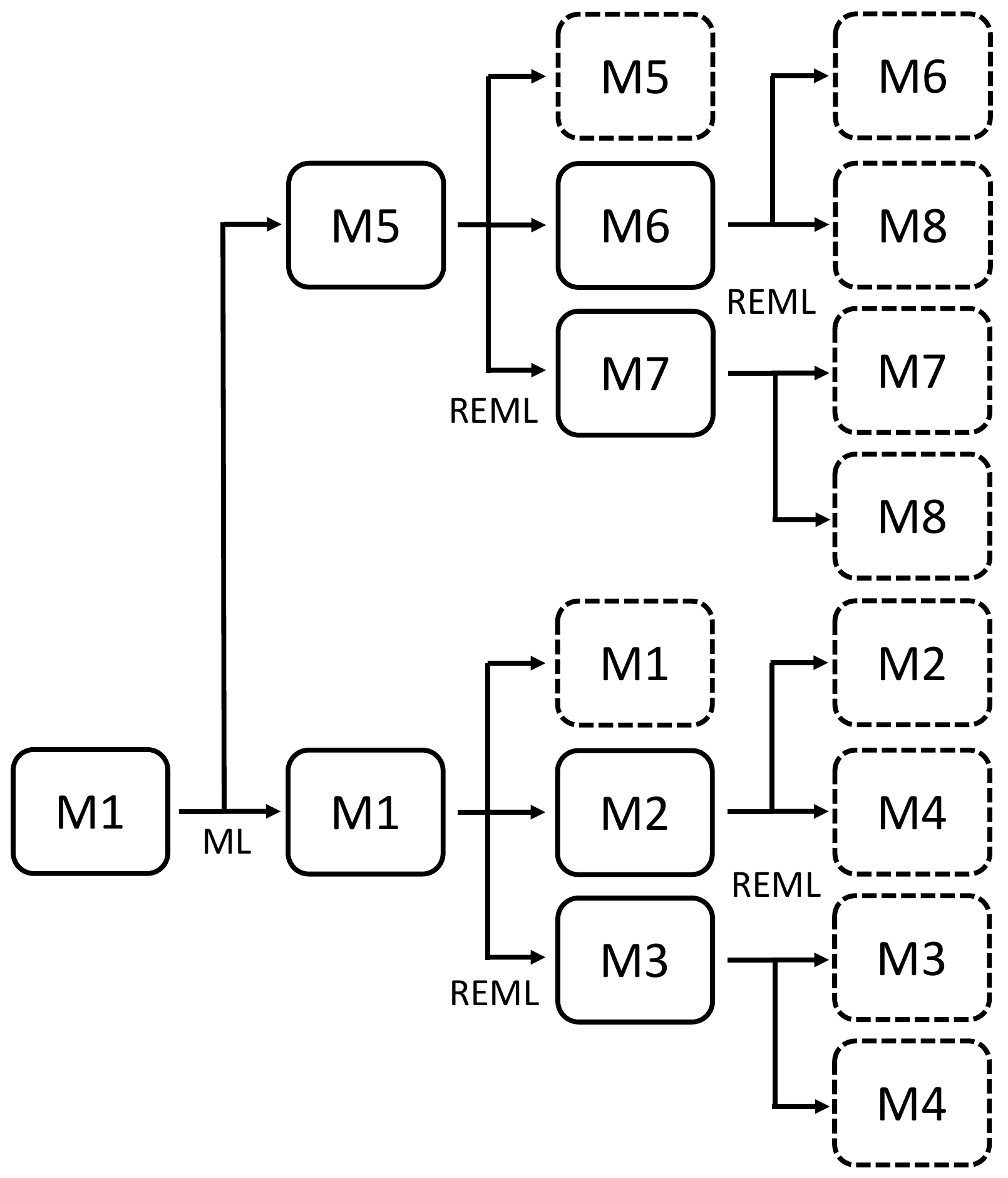}
        \hspace{2em}
        \includegraphics[width=0.40\textwidth]{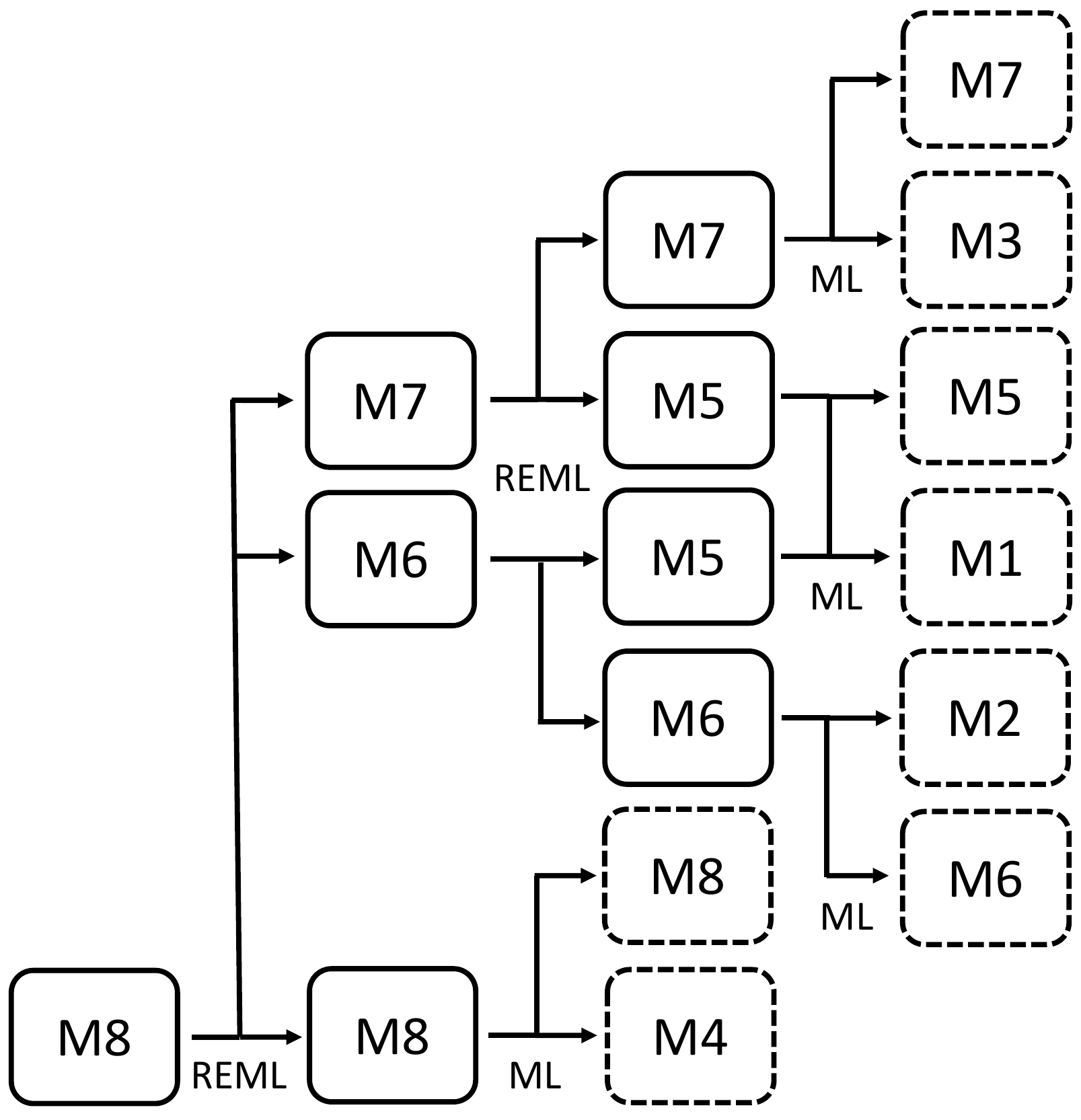}
    \end{center}
    \caption{Diagram depicting the forward (left) and backward (right) stepwise selection procedures. Arrows visualize the selection flow, nodes in wrapped in full line visualize the non-terminal steps, and nodes wrapped in dashed line visualize the terminal steps.}
    \label{fig:model_selection}
\end{figure}

\section*{Appendix B: Sensitivity to alternative prior distributions}
For sensitivity analysis, we used the NIH grant proposal example with two covariates, discussed in Section~\ref{sec:RealDataNIH2}. We re-estimated the models under the two additional prior distributions depicted in Figure~\ref{fig:prior_distributions}: (1) a tighter prior distribution with $\sigma = 0.25$, specifying expectation of smaller differences between the groups and (2) a wider prior distribution with $\sigma = 1$, specifying expectation of larger differences between the groups. Table~\ref{tab:sensitivity_BF} describes the resulting inclusion Bayes factors under each parameterization and Table~\ref{tab:sensitivity_estimates} describes the resulting model-averaged parameter estimates under each parameterization. The results from the middle column are identical to those presented in Section~\ref{sec:RealDataNIH2}.

\begin{table}[h!]
\centering
\caption{Inclusion Bayes Factors for Differences between the Groups for Each Parameter (Rows) across the Different Prior Distribution Specifications.}
\begin{tabular}{lrrr}
  \hline
  Prior distribution setting     & $\sigma = 0.25$ & $\sigma = 0.50$ & $\sigma = 1.00$  \\ 
  \hline
  \multicolumn{4}{l}{Residual variances ($\sigma_{\epsilon}$)}                  \\
  Gender    & 0.96                & 0.55                & 0.31                  \\
  Stage     & $2.00 \times 10^{17}$ & $1.09 \times 10^{17}$ & $7.84 \times 10^{16}$   \\
  \hline
  \multicolumn{4}{l}{Structural variances ($\sigma_{\gamma}$)}                  \\ 
  Gender    &  1.06               & 0.70                & 0.43                  \\ 
  Stage     &  7.80               & 4.74                & 2.60                  \\ 
  \hline
  \multicolumn{4}{l}{Mean ($\mu$)}                                              \\ 
  Gender    & 0.36                & 0.18                & 0.09                  \\ 
  Stage     & $7.77 \times 10^{30}$ & $1.24 \times 10^{31}$ & $9.67 \times 10^{30}$   \\ 
  \hline
\end{tabular}\label{tab:sensitivity_BF}
\end{table}

\begin{table}[h!]
\centering
\caption{Posterior Model-Averaged Estimates (and 95\% Credible Intervals) of Differences/Standard Deviation Ratios between the Groups for Each Parameter (Rows) across the Different Prior Distribution Specifications.}
\begin{tabular}{lrrr}
  \hline
  Prior distribution setting     & $\sigma = 0.25$ & $\sigma = 0.50$ & $\sigma = 1.00$  \\ 
  \hline
  \multicolumn{4}{l}{Residual standard deviation ratio}                  \\
  Gender    & 0.97 (0.91 -- 1.00) & 0.98 (0.91 -- 1.00) & 0.99 (0.92 -- 1.00)\\
  Stage     & 0.78 (0.74 -- 0.83) & 0.78 (0.74 -- 0.83) & 0.78 (0.74 -- 0.83)\\
  \hline
  \multicolumn{4}{l}{Structural standard deviation ratio}                  \\ 
  Gender    &  0.95 (0.81 -- 1.00) & 0.96 (0.81 -- 1.00) & 0.97 (0.81 -- 1.00) \\ 
  Stage     &  0.84 (0.72 -- 1.00) & 0.85 (0.72 -- 1.00) & 0.87 (0.72 -- 1.00) \\ 
  \hline
  \multicolumn{4}{l}{Mean difference}                                              \\ 
  Gender    & -0.01 (-0.10 --  0.00) & -0.01 (-0.08 --  0.00)  & -0.00 (-0.07 --  0.00)) \\ 
  Stage     & -0.55 (-0.63 -- -0.47) & -0.56 (-0.65 -- -0.48) & -0.57 (-0.65 -- -0.48) \\ 
  \hline
\end{tabular}\label{tab:sensitivity_estimates}
\end{table}

As expected, the different prior distributions had a noticeable effect on the degree of evidence either for/against the differences in the means/standard deviation ratios. Table~\ref{tab:sensitivity_BF} suggests more evidence for the null hypotheses with increasing standard deviation of the prior distribution, however, the qualitative conclusions remain essentially unchanged. Evidence against the presence of differences in the residual variances was, at most, between weak and moderate evidence, the later only under a prior distribution expecting large effects. The best performing models (not displayed) are also similar under all model specifications, although the tightest prior distributions ($\sigma = 0.25$) included both gender and stage as a predictor for the structural variances in the model with the highest posterior model probability (0.21) whereas the models with the medium ($\sigma = 0.50$) and wide ($\sigma = 1.00$) prior distributions included only stage for all three parameters in the model with the highest posterior model probability (0.23 and 0.34 respectively). Most importantly, as suggested by Table~\ref{tab:sensitivity_estimates}, the posterior model-averaged estimates remain essentially unchanged under the different prior distribution specifications.

\section*{Appendix C: Mean-unadjusted results for the NIH data}

Here we present results assuming constant overall mean, i.e., restricting $\boldsymbol{\beta}_\mu$ in Equation~\ref{eq:parameterization} to $\boldsymbol{0}$, which corresponds to evaluating variance components and IRR for the covariate unadjusted ratings. The best performing model suggests differences in the structural and also in the residual variance for the two stage groups, but no effect of gender (Table~\ref{tab:NIH_best_models_unadj}).

\begin{table}[ht]
\centering
\caption{
Model Structure for the Best Performing Models in the NIH Data Set when Considering Both the Gender and Career Stage as Predictors for Variance Components, and Assuming Constant Overall Mean. \emph{Note:} First three columns describe the model in terms of predictors of each parameter ($\mu$, $\sigma_\gamma$, and $\sigma_\epsilon$). Marg. Lik denotes the marginal likelihood, $p(\mathcal{M}_{i})$ the prior model probability, and $p(\mathcal{M}_{i} \given \text{data})$ the posterior model probability of each model.}
\label{tab:NIH_best_models_unadj}
\begin{tabular}{cccrrr}
  \hline
  $\mu$ & $\sigma_\gamma$ & $\sigma_\epsilon$ & Marg. Lik. & $p(\mathcal{M}_{i})$ & $p(\mathcal{M}_{i} \given \text{data})$ \\ 
  \hline
  
  None            & Stage           & Stage           & -7743.49 & 0.06 & 0.44 \\ 
  None            & Stage           & Gender \& Stage & -7744.02 & 0.06 & 0.26 \\ 
  None            & Gender \& Stage & Stage           & -7744.11 & 0.06 & 0.24 \\ 
  None            & Gender \& Stage & Gender \& Stage & -7745.41 & 0.06 & 0.06 \\ 
\hline
\end{tabular}
\end{table}

As in the covariate adjusted ratings (Section~\ref{sec:RealDataNIH2}), we find a very strong evidence for the difference between the two career stage groups in residual variance, $\text{BF}_{\epsilon, \bar{\epsilon}} = 9.84 \times 10^{16}$, and we find weak evidence against the difference in residual variance between the two gender groups, $\text{BF}_{\epsilon, \bar{\epsilon}} = \nicefrac{1}{2.09} = 0.48$. The model-averaged estimates of the residual standard deviation ratios are 0.79 (0.75 -- 0.83) for the two career stage groups and 0.98 (95\% central credible interval: 0.92 -- 1.00) for the two gender groups.

However, unlike in the case of covariate adjusted means presented in Section~\ref{sec:RealDataNIH2}, for the structural variance, we find very strong evidence for the difference between the two career stage groups, $\text{BF}_{\gamma, \bar{\gamma}} = 2.99 \times 10^{16}$, 
with the model-averaged posterior mean estimates of the structural standard deviation ratios 
of 0.79 (0.75 -- 0.83) for the two career stage groups.

The high values of structural variance in the Non-Experienced group subsequently lead to higher IRR values, see Table~\ref{tab:NIH_parameters_unadjusted}.
\begin{table}[ht]
\centering
\caption{Estimated Model-Averaged Marginal Means and 95\% CI for Each of the Parameters. \\ \emph{Note:} Exp - Experienced, nExp -- Non-Experienced}
\begin{tabular}{rrrrrr}
  \hline
  Gender & Stage & \multicolumn{1}{c}{$\mu$} & \multicolumn{1}{c}{$\sigma_\gamma$} & \multicolumn{1}{c}{$\sigma_\epsilon$} & \multicolumn{1}{c}{IRR} \\ 
  \hline
  Female & nExp &  -0.06 [-0.09, -0.02] & 0.81 [0.73, 0.91] & 0.97 [0.93, 1.03] & 0.41 [0.36, 0.47] \\ 
  Male & nExp   &   -0.06 [-0.09, -0.02] & 0.79 [0.71, 0.87] & 0.95 [0.91, 0.99] & 0.41 [0.35, 0.46] \\ 
  Female & Exp     & -0.06 [-0.09, -0.02] & 0.53 [0.48, 0.60] & 0.76 [0.73, 0.81] & 0.33 [0.27, 0.39] \\ 
  Male & Exp       &  -0.06 [-0.09, -0.02] & 0.51 [0.47, 0.56] & 0.75 [0.72, 0.78] & 0.32 [0.27, 0.37] \\ 
   \hline
\end{tabular}\label{tab:NIH_parameters_unadjusted}
\end{table}

\end{document}